\pdfoutput=1 
\RequirePackage{fix-cm}
\documentclass[smallextended,final,natbib]{svjour3}       
\smartqed  
\usepackage{natbib} 
\usepackage{epsfig,graphicx,dcolumn,bm,times,amsmath,amssymb}
\usepackage{tabularx}
\usepackage{enumerate}
\usepackage[colorlinks,linkcolor=blue,anchorcolor=green,citecolor=blue]{hyperref}
\newcommand{\mytilde}{\raise.17ex\hbox{$\scriptstyle\mathtt{\sim}$}}

\begin{document}
\bibliographystyle{plainnat}

\title{Prevention of infectious diseases by public vaccination and individual protection
}


\author{Xiao-Long Peng \and
        Xin-Jian Xu \and
        Michael Small \and
        Xinchu Fu \and
        Zhen Jin  
}


\institute{Xiao-Long Peng \at
              Complex Systems Research Center, Shanxi University, Taiyuan, Shanxi 030006, China \\%
              \email{xlpeng@sxu.edu.cn}           
           \and
           Xin-Jian Xu \at
              Department of Mathematics, Shanghai University, Shanghai 200444, China \\
              \email{xinjxu@shu.edu.cn}           
           \and
           Michael Small \at
             School of Mathematics and Statistics, The University of Western Australia, Crawley, WA 6009, Australia \\
             Mineral Resources, CSIRO, Kensington, WA 6151, Australia\\
             \email{michael.small@uwa.edu.au}                        
           \and
           Xinchu Fu \at
             Department of Mathematics, Shanghai University, Shanghai 200444, China \\
             \email{xcfu@shu.edu.cn}           
           \and
           Zhen Jin \at
             Complex Systems Research Center, Shanxi University, Taiyuan, Shanxi 030006, China\\
             \email{jinzhen@sxu.edu.cn}
}

\date{Received: date / Accepted: date}

\maketitle

\begin{abstract}
In the face of serious infectious diseases, governments endeavour to implement containment measures such as public vaccination at a macroscopic level. Meanwhile, individuals tend to protect themselves by avoiding contacts with infections at a microscopic level. However, a comprehensive understanding of how such combined strategy influences epidemic dynamics is still lacking. We study a susceptible-infected-susceptible epidemic model with imperfect vaccination on dynamic contact networks, where the macroscopic intervention is represented by random vaccination of the population and the microscopic protection is characterised by susceptible individuals rewiring contacts from infective neighbours. In particular, the model is formulated both in populations without and then with demographic effects (births, deaths, and migration). Using the pairwise approximation and the probability generating function approach, we investigate both dynamics of the epidemic and the underlying network. For populations without demography, the emerging degree correlations, bistable states, and oscillations demonstrate the combined effects of the public vaccination program and individual protective behavior. Compared to either strategy in isolation, the combination of public vaccination and individual protection is more effective in preventing and controlling the spread of infectious diseases by increasing both the invasion threshold and the persistence threshold. For populations with additional demographic factors, we investigate temporal evolution of infected individuals and infectious contacts, as well as degree distributions of nodes in each class. It is found that the disease spreads faster but is more restricted in scale-free networks than in the Erd\"os-R\'enyi ones. The integration between vaccination intervention and individual rewiring may promote epidemic spreading due to the birth effect. Moreover, the degree distributions of both networks in the steady state is closely related to the degree distribution of newborns, which leads to uncorrelated connectivity. All the results demonstrate the importance of both local protection and global intervention, as well as the demographic effects. Our work thus offers a more comprehensive description of disease containment.
\end{abstract}

\section{Introduction}\label{intro}

Since complex network models \citep{N10} were applied to mathematical epidemiology \citep{KR07} --- in such a way that the host population is modelled as a contact network where nodes represent individuals and links stand for contacts among them --- the topological structure of the underlying network has shown a strong impact on the spread of infectious diseases \citep{LM01,PV01a,MNMS03,KE05,BBV07,MAM09,Durrett10}. For instance, based on mean-field approximations, \citet{PV01b} suggested that there is no typical epidemic threshold in a scale-free (SF) network which follows a power-law degree distribution $P(k)\propto k^{-\gamma}$ with the power exponent $2<\gamma\leq3$ in the thermodynamic limit. Otherwise, there is a positive critical value $\lambda_{\rm c}$ of the transmission rate $\lambda$ only for $\gamma>3$. It is worth remarking that in a more mathematically rigorous analysis, \citet{CD09} proved that the critical value $\lambda_{\rm c}$ is also zero for contact processes on random networks with power-law degree distributions for any power exponent $\gamma>3$. In other words, the infection can pervade the whole SF network provided the transmission rate is nonzero.

Over the past decades, considerable progress has been made in the study of epidemic models on static networks, where the network topology remains fixed during the process of disease transmission. Many topological features of the network, such as the heterogeneity \citep{EK02,Volz08}, degree correlation \citep{BPV03,MGP03} and clustering \citep{M09,VMGM11}, were explicitly addressed, yet most studies ignore the impact of human behavior (in response to infection) on the disease \citep{F07,FGW09,MPAGMV11,PAM12}. Such effect is likely to be especially acute when modifications to human behavior causes changes in the structure of the underlying contact network \citep{GDB06,SS08,ZR08,SSP10}. Most recently, it has been demonstrated that a better understanding of the interplay between human behavior and epidemic dynamics would help design effective intervention strategies such as vaccination and social distancing policies, and improve the cost effectiveness of disease control \citep{FSJ10,FCC11}. It is therefore of great significance to incorporate human behavior in the study of network epidemiology \citep{PBGV11,DC12,WFSX12,BGM13,XL14,WAWWB15}. In general, there are two scenarios of human behavior to be considered in the face of a contagious disease. First, at the macroscopic level authorities enact public containment programs such as quarantine or vaccination; second, at the microscopic level individuals adopt self-protective measures against the disease based on their perceived risk of infection \citep{FSJ10,DC12,M11}.

Vaccination is one of the most effective public policies for preventing the transmission of infectious diseases \citep{BGE03,ABB10}, and to date there has been a wealth of studies on various vaccination strategies for epidemic models in complex networks \citep{TY06,SASB08,JR14}, indicating powerful effects of vaccination on the inhibition of disease transmission, even though some vaccines provide temporary immunity and only reach part of the population. For instance, targeted vaccination towards high-degree nodes is widely known to have a much higher effectiveness than random vaccination --- either in SF networks \citep{PV02} or in small-world networks \citep{ZK02}. However, targeting the highest-degree nodes requires full knowledge of the degree of each node in the network. To overcome this shortcoming, an alternative vaccination strategy based on local information has been proposed \citep{CHB03}. Under this strategy vaccination is implemented on a random acquaintance of a randomly chosen node. Again, this acquaintance vaccination proves to be more effective than random vaccination since it prefers to target high-degree nodes (this is a consequence of the so-called friendship paradox --- on average neighbours of a random node have more neighbours than the node itself).

On the other hand, appropriate and timely response of individuals to a spreading contagion may improve the efficiency of vaccination and save medical resources. Several studies looking at this hypothesis have been framed in network epidemic models through the addition of a rewiring mechanism (for details, see \citealp{GB08,SS10b,JRS13} and the references therein). A susceptible-infected-susceptible (SIS) epidemic model on adaptive networks was first proposed by \citet{GDB06}, where at a given rate susceptible individuals rewire their connections from infected neighbours to the susceptibles. The basic idea of this adaptive network epidemiology is that individuals tend to change their habits and social contacts to avoid infection in response to a disease threat \citep{SS10b}. Still further rewiring scenarios have also been considered reflecting different individual reactions to the infectious status of others. For example, \citet{VM07} introduced a \textit{neighbour exchange} model to account for the impacts of the stability of social contacts on the spread of infectious diseases \citep{REE08}. Alternatively, a random swapping of contact partners between susceptible-infected and infected-infected links was studied in \citet{K10}.

Most adaptive epidemic models have largely focused on the standard SIS model \citep{AM92}, where the rewiring operation isolates infective individuals from the susceptible population in the long run, and therefore influences the epidemic
propagation. In turn, the larger the infected fraction in the population, the more likely susceptible individuals will choose to rewire their contacts with infective neighbours. In other words, the epidemic prevalence can affect individual behavior, while avoiding contacts with infection also exerts a strong impact on epidemic dynamics. Typically, in an adaptive network model both the dynamics of and on the network form a feedback loop \citep{GB08}, which usually involves many interesting dynamical features such as bistability and oscillation \citep{MNHAD10,GDB06,ZR08,JRS13}.

So far, little attention has been paid to the combined effects of public vaccination and individual protection. Only recently have Shaw and Schwartz adopted a random vaccination strategy in an adaptive SIS model to examine the interplay between individual rewiring of social contacts and random vaccination of susceptible individuals \citep{SS10}. In their model, vaccinated individuals secure complete protection from infection unless they return to the susceptible state. That is, the vaccinated can not be infected directly from their infective neighbours. However, there is clear
evidence that some vaccines are not completely efficient, namely, they come with no guarantee of perfect protection from infection \citep{GMNR01,Smith02} --- rather, they confer only partial immunity and a lower transmission rate than the unvaccinated susceptibles. Moreover, in reality, vaccines rarely cover the entire population and only provide finite-time immunity against infection. In light of this, \citet{KV00} proposed a compartmental SIS model with imperfect vaccination, where vaccinated individuals may contract infection directly from their infectious contacts, although at a vaccine-reduced transmission rate \citep{BF04}. Hereafter, we refer to this as the SIV model. Most recently, this imperfect vaccination model has been restudied on static SF networks, revealing surprising epidemic behavior \citep{PXFZ13}. However, the role of individual behavior has been neglected in changing the transmission routes of the disease as well as in reshaping the contact patterns of the underlying network.

This paper aims to provide a systematic framework that couples public vaccination and individual protection. For this purpose, we develop an adaptive SIV model by extending the compartmental SIV model \citep{KV00} to an adaptive network \citep{GDB06}. The adaptive SIV model integrates contact rewiring at the microscopic level with vaccination intervention at the macroscopic level in response to disease propagation. In particular, we consider a population without and with demographic effects, respectively, to account for two different cases as follows: case (1)---the disease spreads through the population much faster than the births and deaths of individuals, and case (2)---the disease spreads on a timescale comparable to the speed of individuals' births and deaths. This observation mainly depends on what type of the disease is being modelled. For instance, the seasonal influenza can be modelled well in a population without demography. Conversely, for the human immunodeficiency virus (HIV) where infection span decades, then the demographic impact should be taken into account \citep{KR07}. We establish a mathematical configuration for both cases, which allows for an in-depth analysis of the mutual interaction between disease propagation and network evolution.

In Sect.~\ref{sec2} we introduce the adaptive SIV model, and construct a set of approximating deterministic equations for both a closed system and one with demography. Section \ref{sec3} contains main results and Sect. \ref{sec4} concludes.

\section{Adaptive SIV Model}
\label{sec2}

\begin{figure}
\begin{center}
\includegraphics[width=0.6\columnwidth]{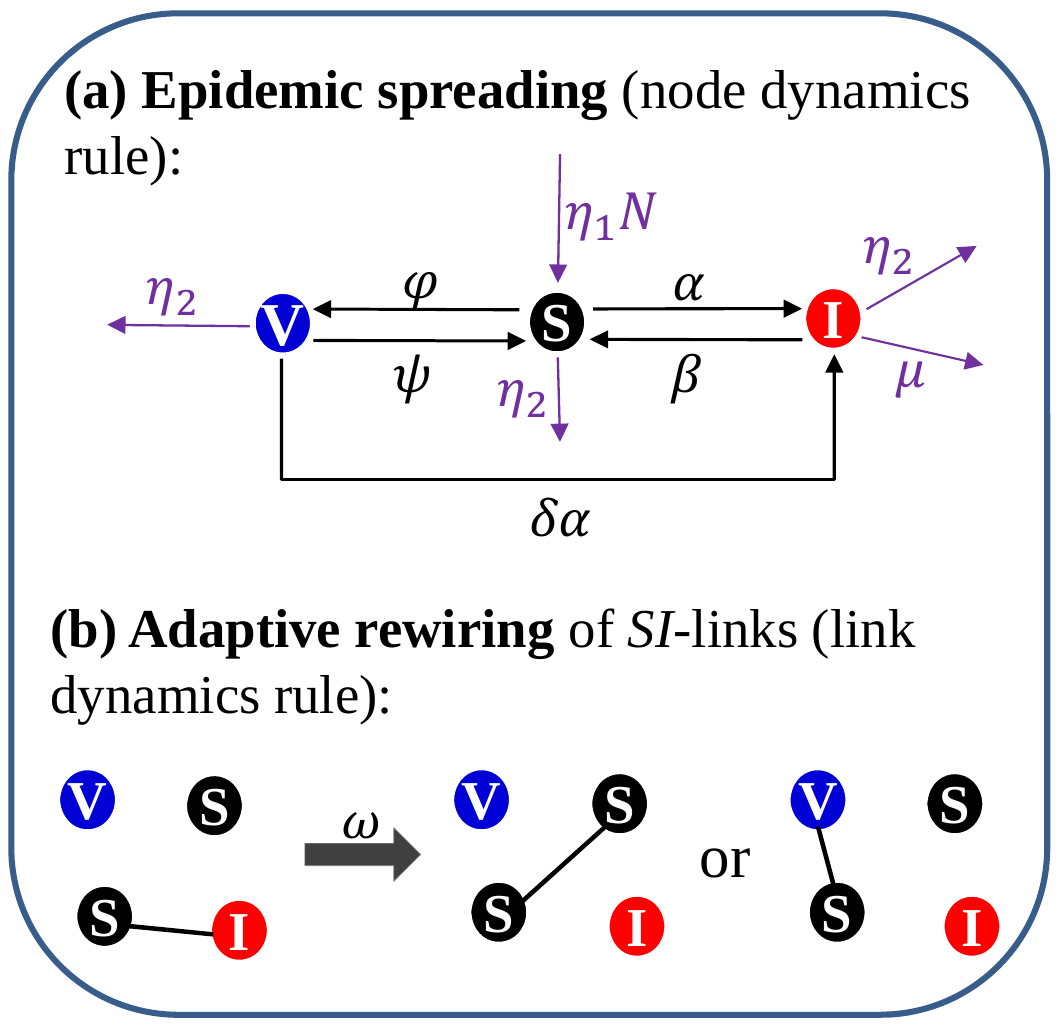}
\caption{(Colour online) Schematic illustration of the adaptive SIV model. The model evolves concurrently according to the node dynamics rule (a) and the link dynamics rule (b) with adaptive rewiring of $SI$-links. At each time step, susceptible nodes are infected by their infected neighbours at a transmission rate $\alpha$ per $SI$-link and are vaccinated at a per capita rate $\varphi$. Infected nodes recover and return to being susceptible at a per capita rate $\beta$. Vaccinated nodes become susceptible at a per capita rate $\psi$ and are infected at a vaccine-reduced transmission rate $\delta\alpha$ per $IV$-link due to imperfect immunity. With the background demographic changes taken into consideration, we assume that each node can give birth to a susceptible node with a birth rate $\eta_1$, and each node dies at a natural death rate $\eta_2$. In addition, the infected nodes die with an extra rate $\mu$ because of infection, namely, each infected node will die at a rate $\eta_2+\mu$. In the adaptive SIV model, we assume that only susceptible nodes are permitted to rewire connections away from their infected neighbours at a rate $\omega$ per $SI$-link. In other words, with rate $\omega$, for each $SI$-link, the susceptible node breaks the link to the infected node and forms a new link to a randomly selected susceptible or vaccinated node.}\label{fig1}\end{center}
\end{figure}

We employ individual rewiring of contacts with infective neighbours as an individualised self-protective behavior and the vaccination scheme (reducing the rate of infection) as a public intervention in reaction to an infectious disease. The nodes of the network represent individuals in the population and links are potentially infectious contacts among them. In the adaptive SIV model, each node may have only one of the three possible states: susceptible ($S$), infected ($I$) and vaccinated ($V$). As demonstrated in Fig.~\ref{fig1}(a), the transition probabilities for node states are as follows: a susceptible node becomes infected with the transmission rate $\alpha$ per $SI$-link (where the parameters in the model definition are listed in Table~\ref{table1});
\begin{table*}[htbp]
 \addtolength{\tabcolsep}{0.5pt}
 \caption{Notation used in the model definition. }\label{table1}
 \hspace{-0.15cm}
 \begin{tabularx}{1\textwidth}{lX}
 \hline
 Term & Meaning\\
 \hline
 $\alpha$ & transmission rate per $SI$-link\\
 $\beta$ & recovery rate from infection\\
 $\varphi$ & vaccination rate of susceptibles\\
 $\psi$ &  relapse rate of vaccinated individuals (i.e., each vaccinated returns to being susceptible after an average time period of $\tau=1/\psi$ due to temporary immunity)\\
 $1-\delta$ & effectivity of the vaccine-induced protection against infection (i.e., vaccinated individuals can be infected at a vaccine-reduced transmission rate $\delta\alpha$)\\
 $\eta_1$ & birth rate\\
 $\eta_2$ & natural death rate\\
 $\mu$ & death rate from infection\\
 $\omega$ & rewiring rate per $SI$-link where the susceptible node breaks away from the infected and then establishes a connection with a randomly chosen noninfected node\\
 \hline
 \end{tabularx}
\end{table*}
an infective node recovers back to the susceptible state with the recovery rate $\beta$; according to random vaccination strategy, a susceptible node is vaccinated at the vaccination rate $\varphi$; a vaccinated node returns to the susceptible state with the resusceptibility rate $\psi$ as the vaccine wears off; due to imperfect immunity, the vaccinated node will be infected directly from one of its infective neighbours with a reduced transmission rate $\delta\alpha$, where the parameter $\delta$ measures the inefficacy of the vaccine-induced protection against infection ($\delta=0$ means the vaccine is completely effective, while $\delta=1$ means the vaccine is totally ineffective). With the background demography taken into consideration, we assume that each node can give birth to a susceptible node with a birth rate $\eta_1$, and each node (irrespective of its state) dies at a natural death rate $\eta_2$. In addition, the infected nodes die with an extra rate $\mu$ because of infection. That is to say, each infected node will die at a rate $\eta_2+\mu$. Meanwhile, the network is rewired adaptively during the disease propagation [see Fig.~\ref{fig1}(b) as an illustration]. For simplicity, only susceptible nodes are allowed to rewire their connections from infected neighbours to randomly selected noninfected (either susceptible or vaccinated) nodes with the rewiring rate $\omega$. Self links and multiple links are prohibited. The adaptive SIV model therefore consists of two processes (Fig.~\ref{fig1}): one is the node dynamics due to the epidemic spreading and the birth and death process, the other is the link dynamics due to the rewiring of $SI$-links of the underlying network.

In what follows, we consider two types of population system --- a closed system without demography and an open system with demography, respectively, to account for the following two different cases: case (1) --- the disease spreads sufficiently fast that the background demographic changes are not significant, and case (2) --- the epidemic occurs on a timescale comparable to the speed of the birth and death process. Therefore, some fast diseases like the seasonal influenza can be well described with the case (1) model, whereas for some slow diseases such as HIV we need to consider the case (2) model.

\begin{table*}[htbp]
 \addtolength{\tabcolsep}{0.5pt}
 \caption{Notation used in the model formulation and analytical approximation. The subscripts $A$, $B$ and $C$ in this table indicate the node state, i.e., $A, B, C\in\{S, I, V\}$, and the lower-case character $a$ ($a\in\{s,i,v\}$) denotes the fraction of nodes in state $A$ ($A\in\{S, I, V\}$) correspondingly.}\label{table2}
 \hspace{-0.15cm}
 \begin{tabularx}{1\textwidth}{lX}
 \hline
 Term & Meaning\\
 \hline
 $A_k$ & number of nodes in state $A$ with degree $k$\\
 $N_{A}=\sum_{k}A_k$ & number of nodes in state $A$\\
 $N_k=\sum_{A}A_k$ & number of nodes with degree $k$\\
 $N=\sum_{k}N_k=\sum_{A}N_{A}$ & total number of nodes in the network \\
 $a={N_{A}}/{N}$ &  fraction of nodes in state $A$, where $a=s, i, v$ corresponding to $A=S, I, V$, respectively\\
 $P(k)={N_k}/{N}$ & degree distribution of the network (i.e., probability that a randomly selected node has a degree $k$)\\
 $\langle k\rangle=\sum_{k}kP(k)=2E/N$ & average degree of the network\\
 $p_{Ak}(t)={A_k}/{N_{A}}$ & probability for a node in state $A$ to have a degree $k$\\
 $g_A(x,t)=\sum_{k}p_{Ak}(t)x^{k}$ & probability generating function of $p_{Ak}(t)$\\
 $g^{\prime}_A(x,t)=\frac{\partial}{\partial{x}}g_A(x,t)$ & partial derivative of function $g_A(x,t)$ with respect to $x$\\
 $g^{\prime\prime}_A(x,t)=\frac{\partial^2}{\partial{x^2}}g_A(x,t)$ & second-order partial derivative of function $g_A(x,t)$ with respect to $x$\\
 ${\langle k_{A}\rangle}=\sum_{k}kp_{Ak}(t)=g_{A}^{\prime}(1,t)$ & average degree of nodes in state $A$\\
 $\bar{p}_k$ & degree distribution of newborns (i.e., probability for a node entering the network to have $k$ links)\\
 $\bar{g}(x)=\sum_{k}\bar{p}_kx^{k}$ & probability generating function of $\bar{p}_k$\\
 $\bar{g}^{\prime}(x)=\frac{\rm d}{{\rm d} {x}}\bar{g}(x)$ & first-order derivative of function $\bar{g}(x)$ with respect to $x$\\
 $AB$-link & a link connecting a node in state $A$ with a node in state $B$\\
 $M_{AB}$ & number of $AB$-links\\
 $M_{A}=\sum_{k}kA_{k}=g_{A}^{\prime}(1,t)N_{A}$ & number of links emanating from nodes in state $A$\\
 $E=(\sum_{A}M_{AA}+\sum_{A,B}M_{AB})/2$ & total number of links in the network\\
 $P_{AB}={M_{AB}}/{E}$ & fraction of $AB$-links in the network\\
 $ABC$-type triple & a triple in which a node in state $A$ is connected to a node in state $B$ who in turn is connected to a node in state $C$\\
 $M_{ABC}$ & number of $ABC$-type triples in the network\\
 \hline
 \end{tabularx}
\end{table*}

\subsection{Case (1): Adaptive SIV model without demography}\label{closed}

In this case, obviously the demographic turnover can be ignored by setting the demographic parameters to be zero, i.e., $\eta_1=\eta_2=\mu=0$. There is neither the addition of new nodes nor the removal of old nodes, and the rewiring mechanism conserves the total number of links among the entire network. Therefore, the numbers $N$ (of nodes) and $E$ (of links) are constant during the model evolution.

Following~\citet{GDB06} and \citet{SS10}, we also develop a mean-field model for the system, which tracks the dynamics of both nodes and links. The notation summarised in Table \ref{table2} allows us to obtain the following set of differential equations for the evolution of the fractions of nodes as follows:
\begin{eqnarray}
\frac{\rm d}{{\rm d}t}s &=& \beta i + \psi v - \varphi s - \alpha \frac{E}{N}P_{SI},\label{eq:1}\\
\frac{\rm d}{{\rm d}t}i &=& \alpha\frac{E}{N}P_{{SI}} + \delta\alpha\frac{E}{N}P_{IV} - \beta i,\label{eq:2}\\
\frac{\rm d}{{\rm d}t}v &=& \varphi s - \psi v - \delta\alpha\frac{E}{N}P_{IV}.\label{eq:3}
\end{eqnarray}
On the right hand side (rhs) of Eq.~(\ref{eq:1}), the first and second terms correspond to the conversion of infected and vaccinated nodes into susceptible ones, respectively, while the third term indicates the vaccination process of susceptible nodes, and the fourth term represents the loss of susceptible nodes due to infection, which is proportional to the number $EP_{SI}$ of $SI$-links and the transmission rate $\alpha$ per $SI$-link. On the rhs of Eq.~(\ref{eq:2}), the first and second terms consider that infection spreads along each $SI$-link with rate $\alpha$ and along each $IV$-link with rate $\delta\alpha$, respectively, while the third term describes recovery of infected nodes at rate $\beta$. Similarly, the first term on the rhs of Eq.~(\ref{eq:3}) describes the gain of vaccinated nodes due to vaccination of susceptible nodes, while the second and third terms describe the loss of vaccinated nodes as a result of immunity wearing off and infection along each $IV$-link, respectively. The nodal dynamics interact closely with the dynamics of links through the change of nodal states and infection of noninfected nodes along $SI$- and $IV$-links. In addition, rewiring of $SI$-links also implicitly affects the system since it alters the numbers of $SI$-, $SS$-, and $SV$-links. To close the model system, we use the low-order moment closure approximation $M_{ABC}\approx M_{AB}M_{BC}/{N_B}$ \citep{Keeling97} with the coefficient $(\langle k\rangle-1)/\langle k\rangle$ being omitted, where $A,B,C\in\{S,I,V\}$ and $M_{ABC}$ represents the number of $ABC$-type triples in which a node in state $A$ is connected to a node in state $B$ who in turn is connected to a node in state $C$. This leads to the following equations for the evolution of the fractions of links:
\begin{eqnarray}
\frac{\rm d}{{\rm d}t} P_{SS}&=& \beta P_{SI} + \psi P_{SV} + \omega\frac{s}{s+v}P_{SI}
 - 2\varphi P_{SS} - 2\alpha \frac{E}{N} \frac{P_{SI}}{s}P_{SS},\label{eq:4}\\
\frac{\rm d}{{\rm d}t} P_{SI}&=& 2\alpha\frac{E}{N}\frac{P_{SS}P_{SI}}{s} + 2\beta P_{II} + (\psi+\delta\alpha\frac{E}{N}\frac{P_{SV}}{v})P_{IV}
- (\alpha+\beta+\varphi+\omega+\alpha\frac{E}{N}\frac{P_{SI}}{s})P_{SI},\quad\quad \label{eq:5}\\
\frac{\rm d}{{\rm d}t}P_{SV} &=& \beta P_{IV} + 2\psi P_{VV} + 2\varphi P_{SS}+\omega\frac{v}{s+v}P_{SI}
- (\alpha\frac{E}{N}\frac{P_{SI}}{s}+\delta\alpha\frac{E}{N}\frac{P_{IV}}{v}+\varphi
+\psi)P_{SV},\label{eq:6}\\
\frac{\rm d}{{\rm d}t} P_{II} &=& \alpha(1+\frac{E}{N}\frac{P_{SI}}{s})P_{SI}+\delta\alpha(1+\frac{E}{N}
\frac{P_{IV}}{v})P_{IV}
 - 2\beta P_{II}, \label{eq:7}\\
\frac{\rm d}{{\rm d}t} P_{IV} &=& \alpha\frac{E}{N}(\frac{P_{SI}P_{SV}}{s}+2\delta\frac{P_{IV}P_{VV}}{v}) + \varphi P_{SI}
 - (\delta\alpha+\delta\alpha\frac{E}{N}\frac{P_{IV}}{v}+\beta+\psi)P_{IV},\label{eq:8}\\
\frac{\rm d}{{\rm d}t} P_{VV} &=& \varphi P_{SV} - 2\psi P_{VV} -2\delta\alpha\frac{E}{N}\frac{P_{IV}P_{VV}}{v},\label{eq:9}
\end{eqnarray}
where the triple approximations in terms of the number of pairs have been absorbed into this formulation. The first three terms on the rhs of Eq.~(\ref{eq:4}) describe the gain of $SS$-links as a result of recovery, loss of immunity, and rewiring, respectively, while the last two terms correspond to the loss of $SS$-links due to vaccination of susceptible nodes and infection through $SSI$-type triples, respectively. Equations (\ref{eq:5})--(\ref{eq:9}) are analogous except that infection of vaccinated nodes at a reduced rate $\delta\alpha$ has been taken into account. Since the numbers $N$ (of nodes) and $E$ (of links) are constant in this model, the whole system is closed under the conservation conditions $s+i+v=1$ and $P_{SS}+P_{SI}+P_{SV}+P_{II}+P_{IV}+P_{VV}=1$. The steady states of the above ordinary differential equations (\ref{eq:1})--(\ref{eq:9}) can be tracked as a function of parameters via the continuation software \cite{AUTO}, which helps investigate the dynamical effects of rewiring and vaccination through the bifurcation analysis (see the Results section \ref{results}).

\subsection{Case (2): Adaptive SIV model with demography}\label{open}

In the previous section, we assume that the time scale of epidemic spread is much faster than the time scale of population births and deaths. This is not adequate for slow diseases which last for decades. To overcome this shortcoming, we take into account the background demographics renewal in the adaptive SIV model, where {(i)} each individual can give birth to a susceptible at a birth rate $\eta_1$ and the newborn randomly connects to $k$ existing individuals according to a preassigned birth degree distribution $\bar{p}_k$, which corresponds to the probability generating function $\bar{g}(x)$, and meanwhile {(ii)} each individual dies from natural causes at a rate $\eta_2$ and the infective dies at an extra rate $\mu$ (possibly, one expects $\mu>0$ due to the disease itself --- something which is omitted from SIS/SIV models without demography). We do not consider the possibility of either vaccination or infection being passed directly to offspring.

The low-dimensional mean-field equations in the case (1) allows us to carry out in-depth bifurcation analysis and predict rich dynamical features for the closed system. However, the case (1) model is based on a homogeneous mixing assumption and low-order moment-closure approximations, which fundamentally neglect the effects of the underlying network structure, generally resulting in inaccurate predictions about the time evolution of the system \citep{MNHAD10}. One important network property to be considered is the heterogeneity of the degree distribution among nodes \citep{PV01a,PV01b}. To make our model applicable to heterogeneous networks, the network nodes are assembled into subgroups $A_{k}$ according to their degree $k$, where $A_{k}$ denotes the number of nodes in state $A\in\{S, I, V\}$ with $k$ connections. With the notation listed in Table~\ref{table2}, one can write the evolution equations for the numbers of susceptible, infected and vaccinated nodes with $k$ connections $S_{k}$, $I_{k}$, and $V_{k}$ as follows:
\begin{eqnarray}
\frac{\rm d}{{\rm d}t}S_{k} & = & \eta_{1}N\bar{p}_{k}-\eta_{2}S_{k}
-\alpha\frac{M_{SI}}{M_{S}}kS_{k}-\varphi S_{k}  +\eta_{1}\bar{g}^\prime(1)(S_{k-1}-S_{k})
 +\beta I_{k}+\psi V_{k} \nonumber \\
 && +\omega M_{SI}\frac{S_{k-1}-S_{k}}{N_{S}+N_{V}}
 -\eta_{2}[kS_{k}-(k+1)S_{k+1}]
 -\mu\frac{M_{SI}}{M_{S}}[kS_{k}-(k+1)S_{k+1}], \label{eq:10} \\
\nonumber &&\\
\frac{\rm d}{{\rm d}t}I_{k} & = & -(\eta_{2}+\mu)I_{k}
 +\alpha\frac{M_{SI}}{M_{S}}kS_{k}+\delta\alpha\frac{M_{IV}}{M_{V}}kV_{k}  +\eta_{1}\bar{g}^\prime(1)(I_{k-1}-I_{k})
  -\beta I_{k} \nonumber \\
  && -\omega \frac{M_{SI}}{M_{I}}[kI_{k}-(k+1)I_{k+1}]  -\eta_{2}[kI_{k}-(k+1)I_{k+1}]
   -\mu\frac{2M_{II}}{M_{I}}[kI_{k}-(k+1)I_{k+1}], \quad\quad\quad \label{eq:11} \\
\nonumber &&\\
\frac{\rm d}{{\rm d}t}V_{k} & = & -\eta_{2}V_{k} +\varphi S_{k} - \delta\alpha\frac{M_{IV}}{M_{V}}kV_{k} +\eta_{1}\bar{g}^\prime(1)(V_{k-1}-V_{k})  -\psi V_{k} \nonumber \\
&& +\omega M_{SI}\frac{V_{k-1}-V_{k}}{N_{S}+N_{V}}  -\eta_{2}[kV_{k}-(k+1)V_{k+1}]
 -\mu\frac{M_{IV}}{M_{V}}[kV_{k}-(k+1)V_{k+1}].\label{eq:12}
\end{eqnarray}
On the rhs of Eq.~(\ref{eq:10}), the first (second) term denotes the creation (loss) of $S_k$ due to natural births (deaths), where $\eta_{1}N$ is the increasing rate of newborn susceptibles that are of degree $k$ with probability $\bar{p}_{k}$ (according to the degree distribution of newborns). The third term corresponds to the loss of $S_k$ because of infection that is proportional to the number of connections $k$, the transmission rate per $SI$-link $\alpha$, and the probability $M_{SI}/M_S$ for a link originating from the susceptible to point to the infective, which is assumed to have no degree correlation with the starting node. The fourth term represents the removal of $S_k$ due to vaccination. The fifth term considers a growth rate $\eta_{1}N$ of newborns, each of which on average has a number $\langle \bar{k}\rangle =\bar{g}^\prime(1) =\sum_{k}k\bar{p}_{k}$ of links. The number $S_k$ increases (decreases) when the susceptibles with degree $k-1$ (degree $k$) are connected to by one of the $\eta_{1}N\bar{g}^\prime(1)$ links emanating from the newborn nodes. In addition, the probability for the newborns to randomly link to the susceptibles with $k$ contacts is $S_{k}/N$. Note that the establishment of connections of the newborns is simply based on the random attachment mechanism. The sixth and seventh terms refer to the recovery process from infection and the relapse to susceptibility for vaccinated nodes as the vaccine wears off, respectively. The eighth term is attributable to the rewiring process with the rewiring rate $\omega$ per $SI$-link. The number $S_k$ increases (decreases) when the susceptibles with degree $k-1$ (degree $k$) is randomly rewired towards by the susceptible endpoint on the $SI$-link after disconnecting from the infected endpoint. And, the probability for the susceptible endpoint to randomly rewire to the susceptibles with degree $k$ from the pool of noninfected nodes is $S_{k}/(N_S+N_V)$. The ninth term depicts the changes of $S_k$ owing to contacts lost from naturally dying nodes. The number $S_k$ increases (decreases) when one of the $k+1$ neighbours ($k$ neighbours) of the susceptible with degree $k+1$ (degree $k$) dies naturally at a rate $\eta_2$. The tenth term describes the changes of $S_k$ due to contacts lost from nodes dying from infection. The number $S_k$ increases (decreases) when one of the infected neighbours of the susceptibles with degree $k+1$ (degree $k$) dies from infection at a rate $\mu$. There is a fraction $M_{SI}/M_S$ of infected nodes among the $kS_k$ neighbours of the susceptibles with degree $k$. Here, we assume that the probability for a link emanating from a susceptible node to point to an infected node is uncorrelated to the degree $k$ of the starting node. The derivation of Eqs.~(\ref{eq:11}) and (\ref{eq:12}) can be explained analogously. The evolution equations for the numbers of susceptible ($N_{S}$), infected ($N_{I}$), and vaccinated ($N_{V}$) nodes, as well as the numbers of links emanating from susceptible ($M_S$), infected ($M_I$), and vaccinated ($M_V$) nodes, can be derived according to Eqs.~(\ref{eq:10})--(\ref{eq:12}), as given in detail in Appendix A.

To close the system and to take into account the dynamical correlations in the infection status of connected individuals, additional equations for the evolution of the number $M_{AB}$ of $AB$-links are required, where $A,B\in\{S,I,V\}$. In a similar manner, we obtain
\begin{eqnarray}
\frac{\rm d}{{\rm d}t}M_{SS} &=& \eta_{1}\bar{g}^{\prime}(1)N_{S}+\beta M_{SI}
+\psi M_{SV}+\omega M_{SI}\frac{N_S}{N_S+N_V} -2\eta_{2} M_{SS}-2\varphi M_{SS}\nonumber \\
&&-\alpha M_{SI}\frac{2M_{SS}}{M_S}
\frac{g_{S}^{\prime\prime}(1,t)}{g_{S}^{\prime}(1,t)},\label{eq:13}\\
\nonumber &&\\
\frac{\rm d}{{\rm d}t}M_{SI}&=& \eta_{1}\bar{g}^{\prime}(1)N_I
+2\beta M_{II}+\psi M_{VI}+\alpha M_{SI}\frac{2M_{SS}}{M_S}
\frac{g_{S}^{\prime\prime}(1,t)}{g_{S}^{\prime}(1,t)}
+\delta\alpha M_{IV}\frac{M_{SV}}{M_V}\frac{g_{V}^{\prime\prime}(1,t)}
{g_{V}^{\prime}(1,t)}\nonumber\\
&& -(2\eta_2+\mu+\alpha+\beta+\varphi+\omega)M_{SI}
-\alpha M_{SI}\frac{M_{SI}}{M_S}\frac{g_{S}^{\prime\prime}(1,t)}
{g_{S}^{\prime}(1,t)},\label{eq:14}\\
\nonumber &&\\
\frac{\rm d}{{\rm d}t}M_{SV} &=& \eta_{1}\bar{g}^{\prime}(1)N_V+\beta M_{IV}
+2\varphi M_{SS}+2\psi M_{VV}+\omega M_{SI}\frac{N_V}{N_S+N_V}-(2\eta_{2}+\varphi+\psi)M_{SV}\nonumber\\
&&-\alpha M_{SI}\frac{M_{SV}}{M_S}
\frac{g_{S}^{\prime\prime}(1,t)}{g_{S}^{\prime}(1,t)}-\delta\alpha M_{IV}\frac{M_{SV}}{M_V}
\frac{g_{V}^{\prime\prime}(1,t)}{g_{V}^{\prime}(1,t)},\label{eq:15}\\
\nonumber &&\\
\frac{\rm d}{{\rm d}t}M_{II} &=& \alpha M_{SI}+\delta\alpha M_{IV}+
\alpha M_{SI}\frac{M_{SI}}{M_S}\frac{g_{S}^{\prime\prime}(1,t)}
{g_{S}^{\prime}(1,t)}+\delta\alpha M_{IV}\frac{M_{IV}}{M_V}\frac{g_{V}^{\prime\prime}(1,t)}{g_{V}^{\prime}(1,t)}
-2(\eta_2+\mu+\beta)M_{II},\quad\quad\quad \label{eq:16}\\
\nonumber &&\\
\frac{\rm d}{{\rm d}t}M_{IV} &=& \varphi M_{SI}+\alpha M_{SI}\frac{M_{SV}}{M_S}\frac{g_{S}^{\prime\prime}(1,t)}{g_{S}^{\prime}(1,t)}
+\delta\alpha M_{IV}\frac{2M_{VV}}{M_V}\frac{g_{V}^{\prime\prime}(1,t)}{g_{V}^{\prime}(1,t)}
-(2\eta_{2}+\mu+\delta\alpha+\beta+\psi)M_{IV}\nonumber\\
&&-\delta\alpha M_{IV}\frac{M_{IV}}{M_V}\frac{g_{V}^{\prime\prime}(1,t)}
{g_{V}^{\prime}(1,t)},\label{eq:17}\\
\nonumber &&\\
\frac{\rm d}{{\rm d}t}M_{VV} &=& \varphi M_{SV}-2(\eta_{2}+\psi)M_{VV}-\delta\alpha M_{IV}
\frac{2M_{VV}}{M_V}\frac{g_{V}^{\prime\prime}(1,t)}{g_{V}^{\prime}(1,t)}.\label{eq:18}
\end{eqnarray}
On the rhs of Eq.~(\ref{eq:13}), the first four terms describe the addition of $SS$-links as a result of newborn nodes (with an average connectivity $\langle \bar{k}\rangle=\bar{g}^{\prime}(1)$) connecting to old susceptible ones with probability $N_S/N$, the recovery of the infected endpoints on $SI$-links at rate $\beta$, the resusceptibility of vaccinated endpoints of $SV$-link due to temporary immunity, and, the rewiring of $SI$-links away from the infected nodes and towards other susceptible ones with probability $\omega{N_S}/{(N_S+N_V)}$, respectively; while the last three terms correspond to the loss of $SS$-links due to the natural death (at a rate $2\eta_2$) and the vaccination (at a rate $2\varphi$) of one of the two endpoints of $SS$-links, and the infection (at a rate $\alpha$) of the intermediate susceptible node along the $SSI$-type triples, respectively. The appearance of the probability generating functions in Eq.~(\ref{eq:13}) represents an improvement in the approximations of the number of triples in terms of the number of pairs by taking into account the heterogeneity of the central node's degree $k$, rather than simply assuming a homogeneous mixing. A complete description of this type of formulation is given in Appendix B. Explanations for the derivation of Eqs.~(\ref{eq:14})--(\ref{eq:18}) are similar except that the infection of vaccinated nodes at a vaccine-reduced transmission rate $\delta\alpha$ has also been considered.

Equations (\ref{eq:13})--(\ref{eq:18}) characterise the contact behavior among individuals of different states and how they change in response to epidemic transmission and adaptive rewiring of $SI$-links. In conjunction with the set of Eqs.~(\ref{eq:10})--(\ref{eq:12}), they govern the temporal evolution of both the epidemic and the underlying contact network. Our analytic approach in Eqs.~(\ref{eq:10})--(\ref{eq:18}) improves on the usual heterogeneous mean-field approximation by \cite{PV01a,PV01b} as we take into account both the connectivity heterogeneity but also the dynamical correlation among the infection states of adjacent nodes, which is generally ignored in the heterogeneous mean-field approximation \citep{GMWPM12}. It is worth noting that \cite{MNHAD10} has recently proposed an improved heterogeneous moment-closure approximation, referred to as the active neighbourhood approach, to study the SIS epidemic dynamics on adaptive networks. This active neighbourhood approach categorises all the nodes based on the number of their susceptible and infected neighbours. In the approximation for the number of triples, our approach only takes the central node's degree into consideration, but the active neighbourhood approach by \cite{MNHAD10}, by contrast, takes the starting node's and the central node's degrees (both the susceptible degree and infectious degree) into account. In this regard, our approach may lose some accuracy for the approximation of number of triples. On the basis of the active neighbourhood approach, \cite{SDHG14} used generating functions with two spatial variables to investigate an adaptive voter model, providing new insights into the connection between network science and the theory of partial differential equations. Moreover, under the formalism framework of the active neighbourhood approximation approach, some improved moment-closure approximations have also been discussed for adaptive networks\citep{SS13,DVBG14}. Due to the detailed information about the total degree and the infectious degree included in the active neighbourhood approach, the above moment-closure approximations are more accurate but is of much greater computation complexity in comparison with the moment-closure approach as in Appendix B, where the approximation of the number of triples only considers the central node's degree. Moreover, all these approximation approaches have assumed zero degree correlation among nodes in the network.

\section{Results}\label{results}\label{sec3}
\subsection{The model without demography}\label{ssec.a}

For the adaptive SIV model without population births and deaths, we focus on the effects of rewiring and vaccination on both dynamics of the epidemic and the network. The steady state of the system of Eqs.~(\ref{eq:1})--(\ref{eq:9})
can be tracked as a function of parameters with the bifurcation software \cite{AUTO}.

\begin{figure}
\begin{center}
\includegraphics[width=\columnwidth]{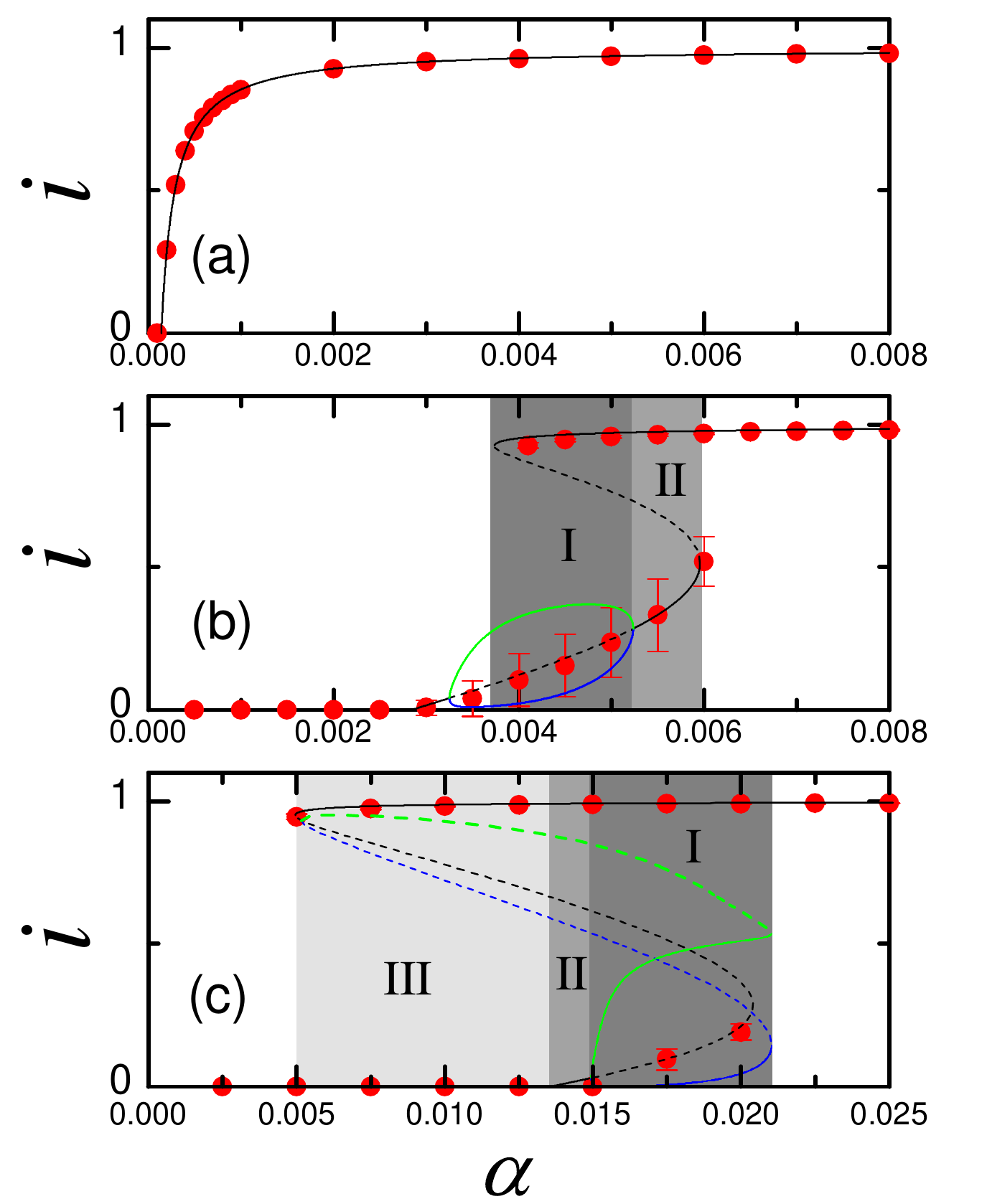}
\caption{(Colour online) Densities of infected nodes $i$ as a function of the transmission rate $\alpha$ for different rewiring rates: $\omega=0$ (a), $0.04$ (b), and $0.2$ (c), respectively. Black solid (dashed) lines: stable (unstable) branches; blue (green) solid lines: lower (upper) turning point of the stable epidemic cycles; blue (green) dashed lines: lower (upper) turning point of the unstable epidemic cycles. Dots and error bars (if larger than the dots): simulation results. All the simulation points are obtained by averaging over 50 runs, each of which goes through $5\times 10^4$ time steps, starting with an initial fraction $0.1\%$ of infected seeds and all the others susceptible. Parameters are $N=10^4$, $E=10^5$, $\beta=0.002$, $\varphi=0.00008$, $\psi=0.0002$, and $\delta=0.0002$. }\label{fig2}\end{center}
\end{figure}

\subsubsection{Bifurcation structure}

We first consider the bifurcation structure of the model. Figure \ref{fig2} shows the bifurcation diagram of the infective fraction $i$ as a function of the transmission rate $\alpha$. The analytic results from the mean-field Eqs.~(\ref{eq:1})--(\ref{eq:9}) are compared with computational simulations of the full model. In the case of no rewiring ($\omega=0$) [Fig.~\ref{fig2}(a)], a single transcritical bifurcation occurs at the well-known epidemic threshold \citep{AM92}, $\alpha_{\rm c}$, which indicates the critical value of transmission rate for invasion of new infections. Hereafter we refer to this threshold as the ``invasion threshold". In this setting, $\alpha_{\rm c}=0.00014$, coincides with the critical behavior $\alpha_{\rm c}={\beta}{(\varphi+\psi)}/({\langle k\rangle}{(\delta\varphi+\psi)})$ obtained in \cite{PXFZ13}. If $\alpha>\alpha_{\rm c}$, the disease-free (healthy) state loses stability while the endemic state becomes stable. As a
low rewiring rate $\omega=0.04$ is introduced [Fig.~\ref{fig2}(b)], the epidemic threshold grows rapidly to $0.00284$. With an increment in $\alpha$, a Hopf bifurcation takes place at $\alpha=0.00325$ with the appearance of a stable epidemic cycle \citep{GDB06} (i.e., the limit cycle, which is a periodic solution), and then a region of periodic oscillations is observed in the range of $0.00325<\alpha<0.00523$. In addition, two saddle-node bifurcations occur at $\alpha=0.00374$ and $\alpha=0.00596$, respectively, between which a clear region of bistability is established,
as shown in the dark gray region \textrm{I} (where a stable endemic state and a stable epidemic cycle coexist) and the medium gray region \textrm{II} (with two coexisting stable endemic states). As the rewiring rate $\omega$ rises to 0.2 [Fig.~\ref{fig2}(c)], the saddle-node bifurcation on the left induces an endemic threshold, above which an already established endemic persists. For convenience, we refer to this threshold as the ``persistence threshold". Across the turning point of the right saddle-node bifurcation (a little to the right), there is a cycle fold bifurcation (i.e., the fold bifurcation of limit cycles) at $\alpha=0.02102$, where stable and unstable epidemic cycles converge, and then disappear. Between the left saddle-node bifurcation and the cycle fold bifurcation, a wider bistable region is located: in the dark gray region \textrm{I}, a stable endemic state and a stable epidemic cycle coexist; in the medium gray region \textrm{II}, there are two stable endemic states; and in the light gray region \textrm{III}, a stable disease-free state coexists with a stable endemic state. Again, an oscillation phenomenon is observed in the dark gray region I.

In our simulations searching for the branch points in the bistable regions in Fig.~\ref{fig2}, we have first examined which attractor the system goes to by the continuation software \cite{AUTO}, thereby we obtained distinctive ranges of different stable branches, and then averaged over considerable runs that appeared to be in the same basin of attraction. In a similar manner to \cite{SS08}, we located the stable endemic state branch by sweeping the transmission rate $\alpha$ from large to small values, and the steady state was averaged over 50 runs, each of which goes through $5\times 10^4$ time steps, starting with an initial fraction $0.1\%$ of infected seeds. The simulation results that dropped into the range of high stable endemic branch were regarded as the high endemic branch point, while the simulation results that dropped into the range of low stable endemic branch were regarded as the low endemic branch point. To locate the stable disease-free state branch, we swept $\alpha$ from zero to large. We considered the disease-free state to be stable if the disease died out in any run of simulation. However, our method might fail to find stable states with small basins of attraction due to the stochastic nature of computational simulations and because the disease-free state is absorbing. Generally speaking, the simulation and the analytic results in Fig.~\ref{fig2} are in good agreement, especially in the upper branches (endemic state) and the lower branches (disease-free state). However, in the oscillatory regions where the limit cycles coexist with a stable endemic state, there are a little discrepancies between theory and simulations. Elsewhere it has been found that there is often disagreement between mean-field (or pairwise) approximation approaches and network simulations about the parameter ranges for oscillations. In Fig.~\ref{fig2}(b), while we have observed periodic oscillations (limit cycles) in the region \textrm{I}, the error bars suggest there is potential for oscillations to appear also in the region \textrm{II}. On the other hand, the error bars in Fig.~\ref{fig2}(c) are so small that it is unclear whether oscillations appear in region \textrm{I} as predicted by the mean-field approximation.

\begin{figure}
\begin{center}
\includegraphics[width=\columnwidth]{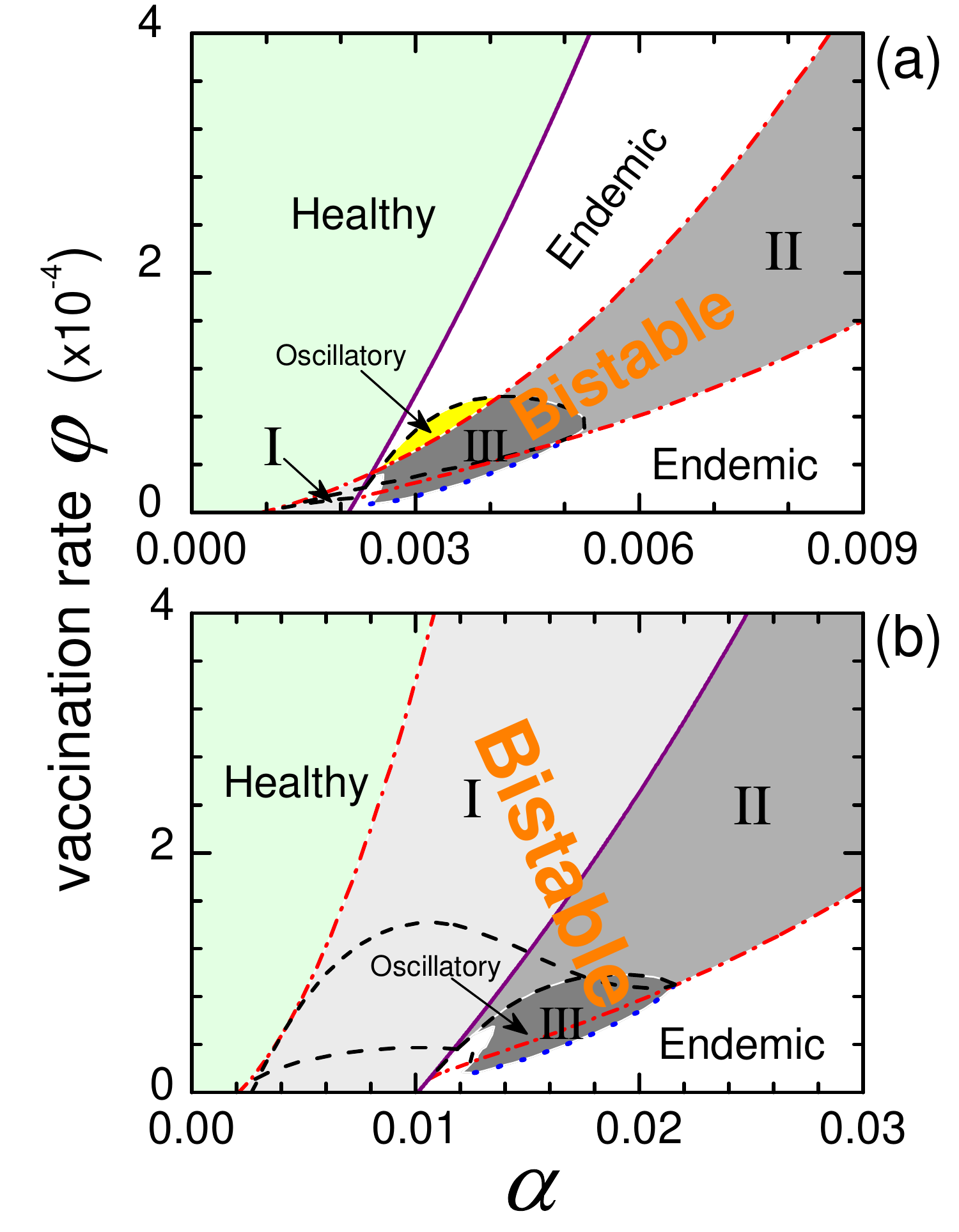}
\caption{(Colour online) The two-parameter bifurcation diagram on the $(\alpha, \varphi)$ plane for different rewiring rates: $\omega=0.04$ (a) and $0.2$ (b). The yellow (in panel (a)), the white and light green regions, respectively, represent a single attractor region, where it is an epidemic cycle in the yellow region (in panel (a)), an endemic state in the white (Endemic) region, and a disease-free state in the light green (Healthy) region. The remaining area represents a bistable region, which is divided into three colours in different gray levels: in the light gray region I both the disease-free state and an endemic state are stable; in the medium gray region II there are two stable endemic states; and in the dark gray region III an endemic state and an epidemic cycle are both stable. Oscillations can be observed both in the yellow region (in panel (a)) and in the dark gray region III due to the presence of a stable epidemic cycle. Purple (solid) lines:
the transcritical bifurcation points; red (dash-dotted) lines: the
saddle-node bifurcation points; black (dashed) lines: the Hopf bifurcation
points; blue (dotted) lines: the cycle fold bifurcation points. Parameters are $N=10^4$, $E=10^5$,
$\beta=0.002$, $\psi=0.0002$, and
$\delta=0.0002$.}\label{fig3}\end{center}
\end{figure}

By varying the vaccination rate $\varphi$ and tracking each bifurcation curve for the rewiring rate $\omega=0.04$, we obtain the two-parameter bifurcation diagram on the $(\alpha, \varphi)$ plane in Fig.~\ref{fig3}(a). In the light green, white, and yellow regions, there is only a single attractor, which is a disease-free state in the light green (Healthy) region, an endemic state in the white (Endemic) region, and an epidemic cycle in the yellow region. The rest of the plane in Fig.~\ref{fig3}(a) represents a bistable region, which is divided into three colours in different gray levels: in the light gray region I both the disease-free equilibrium and an endemic equilibrium are stable; in the medium gray region II there are two stable endemic equilibria; and in the dark gray region III an endemic equilibrium and an epidemic cycle are both stable. Oscillations can be observed both in the yellow region and the dark gray region III since there exists a stable epidemic cycle.

Analogously, we get in Fig.~\ref{fig3}(b) the two-parametric bifurcation diagram for a larger rewiring rate $\omega=0.2$. In the light green (Healthy) region, there is only a stable disease-free equilibrium. Furthermore, there are three different regions of bistability: in the light gray region \textrm{I}, both the disease-free and endemic states are stable; in the medium gray region \textrm{II}, there are two stable endemic states; and in the dark gray region \textrm{III}, a stable epidemic cycle coexists with a stable endemic state. Also, oscillatory dynamics can be observed in the dark gray region \textrm{III}. On the contrary, in the white (Endemic) region, there is only one stable endemic state.

From Fig.~\ref{fig3}, it can be found that for $\varphi=0$ (i.e., without vaccination), there is a backward bifurcation and a saddle-node bifurcation. As the vaccination rate $\varphi$ gradually increases, the model exhibits
a more complicated bifurcation structure with the occurrence of two saddle-node bifurcations, Hopf bifurcations, oscillations, and the cycle fold bifurcation. Finally, for $\varphi$ sufficiently large, the epidemic cycles disappear and the dynamical system is characterised by a transcritical bifurcation and two saddle-node bifurcations. Moreover, the bifurcation points of both the transcritical and saddle-node bifurcations shift rightwards with the increase of $\varphi$. In other words, the public vaccination intervention plays a crucial part in preventing and controlling the spread of infectious diseases by increasing the invasion threshold and also the persistence threshold for epidemics. Furthermore, in comparison with Fig.~\ref{fig3}(a), the disease-free region is larger in Fig.~\ref{fig3}(b), which suggests that the larger the adaptive rewiring rate $\omega$ is, the safer the population is from infection.

\begin{figure}
\begin{center}
\includegraphics[width=\columnwidth]{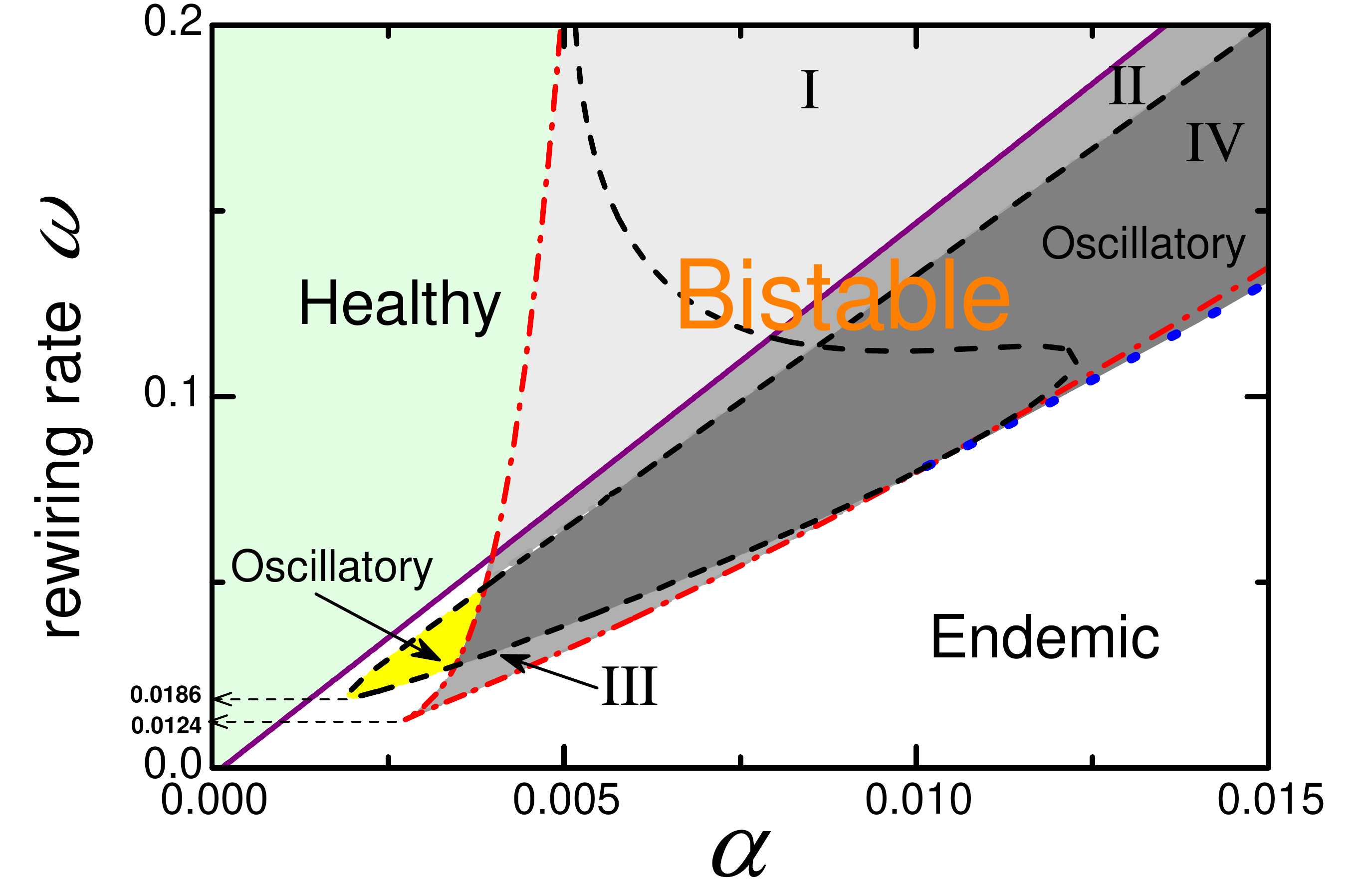}
\caption{(Colour online) Two-parametric bifurcation diagram on the $(\alpha, \omega$) plane. In the white, yellow, and light green regions, there is only a single attractor, which is an endemic state in the white (Endemic) region, an epidemic cycle in the yellow region, and a disease-free state in the light green (Healthy) region. The remaining part
represents a bistable region, which is divided into three colors in different gray levels: In the light gray region \textrm{I} both of the disease-free and endemic states are stable; in the medium gray regions \textrm{II} and \textrm{III} there are two stable endemic states; and in the dark gray region \textrm{IV} a stable endemic equilibrium and a stable epidemic cycle coexist. The oscillatory dynamics can be observed in both the yellow and the dark gray regions since there exists a stable epidemic cycle. The purple solid, red dash-dotted, black dashed, and blue dotted lines, respectively, correspond to the transcritical, saddle-node, Hopf, and cycle fold bifurcation points. Parameters are as in Fig.~\ref{fig2}.}
\label{fig4}\end{center}
\end{figure}

To understand the effect of individual rewiring on the epidemic dynamics in more detail, a two-parameter bifurcation diagram on the plane of $(\alpha, \omega)$ is plotted in Fig.~\ref{fig4}. In the case of no rewiring (i.e., $\omega=0$), there is only the transcritical bifurcation (denoted by a solid line), which separates the healthy and endemic regions. As $\omega$ increases, the system first enters a bistability region (at $\omega=0.0124$), i.e., the medium gray region \textrm{III} (where two stable endemic states coexist), and then undergoes oscillatory dynamics (at $\omega=0.0186$) with the presence of a stable epidemic cycle associated with a supercritical Hopf bifurcation (see the yellow region). As $\omega$ grows further, the system enters another bistable region with one stable endemic state and one stable epidemic cycle (see the dark gray region \textrm{IV}), where the oscillatory phenomena can still be observed. For large enough $\omega$, the whole bistable region is divided by transcritical and Hopf bifurcations into three parts: in the light gray region \textrm{I}, one stable healthy state coexists with one stable endemic state; in the medium gray region \textrm{II}, two stable endemic states coexist; and in the dark gray region \textrm{IV}, one stable endemic state coexists with one stable epidemic cycle. While the increase of $\omega$ slightly enlarges the healthy region, it greatly narrows the endemic region and widens the bistable region. Moreover, the significant role of rewiring in hindering epidemics is clear in the enlarged invasion threshold as well as the emergence of a persistence threshold. According to Figs.~\ref{fig3} and \ref{fig4}, we conclude that the combination of public vaccination and local rewiring is more effective in preventing and controlling the spread of infectious diseases, although it does cause complicated (and interesting) dynamical behavior to occur.

\begin{figure}
\begin{center}
\includegraphics[width=\columnwidth]{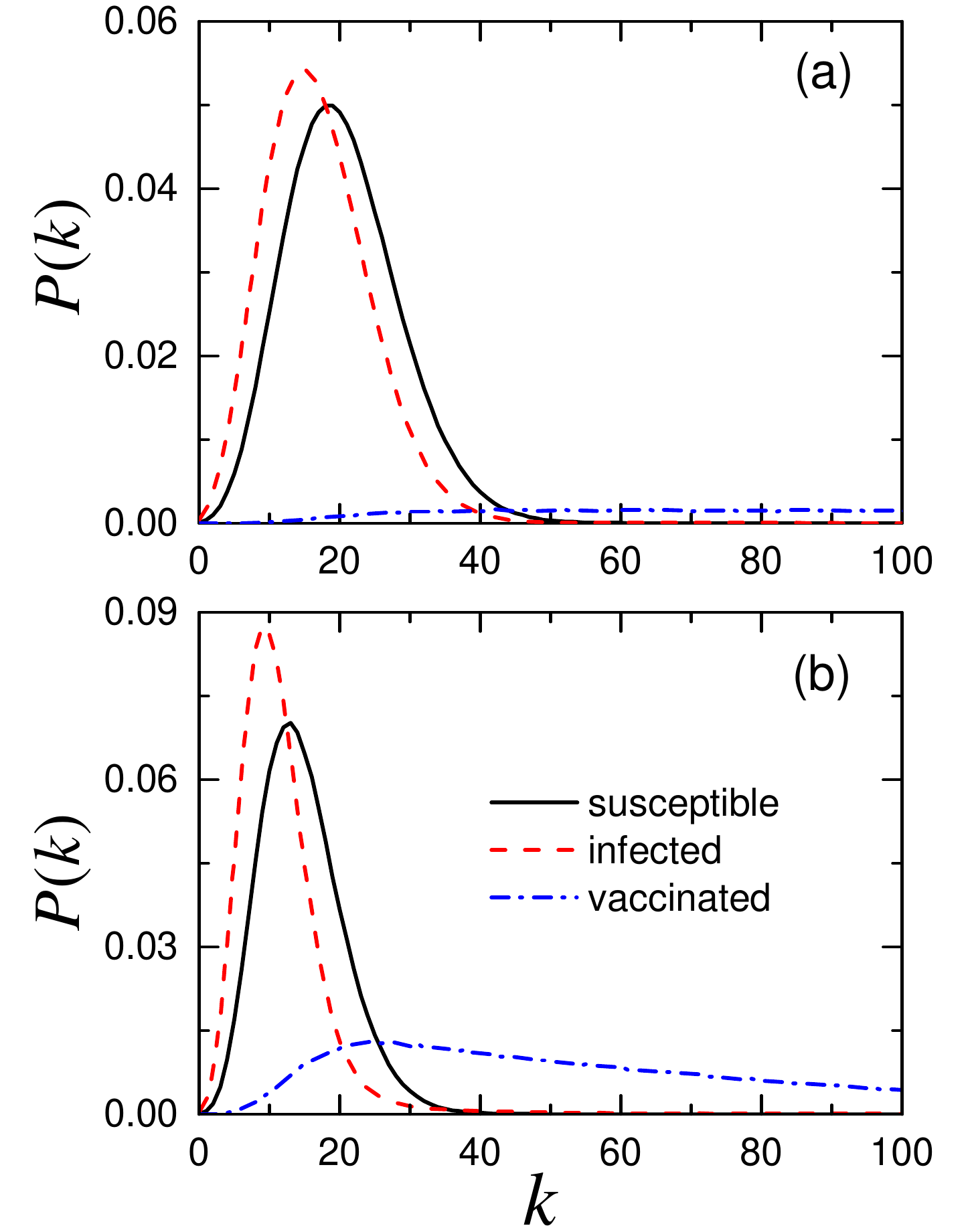}
\caption{(Colour online) Degree distribution in each class of nodes for different epidemic prevalences in the steady endemic state: a high epidemic prevalence $i=0.95$ with average degrees of susceptible, infected, and vaccinated individuals $\langle k_{S}\rangle=21$, $\langle k_{I}\rangle=19$, and $\langle k_{V}\rangle=486$ (a); and a low epidemic prevalence $i=0.53$ with $\langle k_{S}\rangle=15$, $\langle k_{I}\rangle=12$, and $\langle k_{V}\rangle=70$ (b). The black (solid), red (dashed), and blue (dash-dotted) lines correspond to the susceptible, infected, and vaccinated individuals, respectively. The vaccinated individuals have a very broad degree distribution, which extends beyond the domain of the figure. Parameters are $N=10^4$, $E=10^5$, $\alpha=0.006$, $\beta=0.002$, $\varphi=0.00008$, $\psi=0.0002$, $\delta=0.0002$, and $\omega=0.04$.}
\label{fig5}\end{center}
\end{figure}

\subsubsection{Network topology}

We first examine the degree distribution in the endemic steady state. We choose parameters in the bistable region to compare the degree distributions of nodes in each class at different endemic states. In particular, the degree distributions at a high epidemic prevalence $i=0.95$ and a low prevalence $i=0.53$ are shown in Figs.~\ref{fig5}(a) and \ref{fig5}(b), respectively. The results are obtained by averaging over $3\times10^4$ time steps at the steady state with $50$ independent simulations for the model. As shown in Fig.~\ref{fig5}, the average degree of infected nodes is smaller than that of susceptible nodes and far smaller than that of vaccinated ones --- because of rewiring. In the closed population without demographic turnover, the breaking of \textit{SI}-links causes the infective nodes to lose susceptible neighbours, while the susceptible and vaccinated nodes correspondingly gain more connections. Therefore, the average degree of susceptible nodes is larger than the average degree of infected nodes in the steady state. In this regard, the random vaccination of the susceptibles favors higher-degree nodes in the network due to rewiring \citep{SS10}, and thus is expected to be effective, as in previous studies on targeted vaccination \citep{PV02,ZK02}. Since the vaccine-reduced transmission rate is relatively low, vaccinated nodes accumulate connections from rewiring, therefore they have a broad degree distribution with a very large average degree [see $\langle k_{V}\rangle=486$ in Fig.~\ref{fig5}(a) and $\langle k_{V}\rangle=70$ in Fig.~\ref{fig5}(b)]. There are more \textit{SI}-links (at a lower level of epidemic prevalence but with a higher density of susceptible nodes) in Fig.~\ref{fig5}(b) than in Fig.~\ref{fig5}(a), thus adaptive rewiring has stronger effect on the average degree of infected nodes, which in turn reduces the average degree of susceptible nodes due to the recovery process.

It is worth remarking that the presence of the low infection and high infection states is interesting, because the original adaptive network model by \cite{GDB06} just found bistability between endemic and disease-free states. In essence, the differences between the low infection and high infection states are rooted in the different selections of target nodes towards which the original susceptible nodes will be rewired. According to the rewiring mechanism in our model, after susceptible nodes break away their connections from infected neighbours, the susceptibles establish new connections towards randomly selected noninfected nodes, namely, either susceptible or vaccinated. Note that susceptible nodes can be infected at a much higher rate than vaccinated nodes. If vaccinated nodes are selected as the rewiring target, then more vaccinated nodes are able to accumulate connection from rewiring. Thus the mean degree of vaccinated nodes are much higher and they become hub nodes in the network. Among the neighbours of vaccinated nodes, susceptible nodes are in the majority since they are rewired towards the vaccinated nodes. On the other hand, vaccinated nodes are not permitted to rewire away from infected neighbours. Once the vaccinated nodes are infected by one of their neighbours, these hub nodes will induce more infections, leading to a high prevalence in the steady state. On the contrary, if more susceptible nodes are selected as the rewiring target, then the susceptible nodes tend to form a cluster and the vaccinated nodes to some extent lose the ability of accumulating new connections from rewiring. However, susceptible nodes are less likely to become hubs than vaccinated nodes. In this case, the system will end at a low prevalence at the steady state.

\begin{figure}
\begin{center}
\includegraphics[width=\columnwidth]{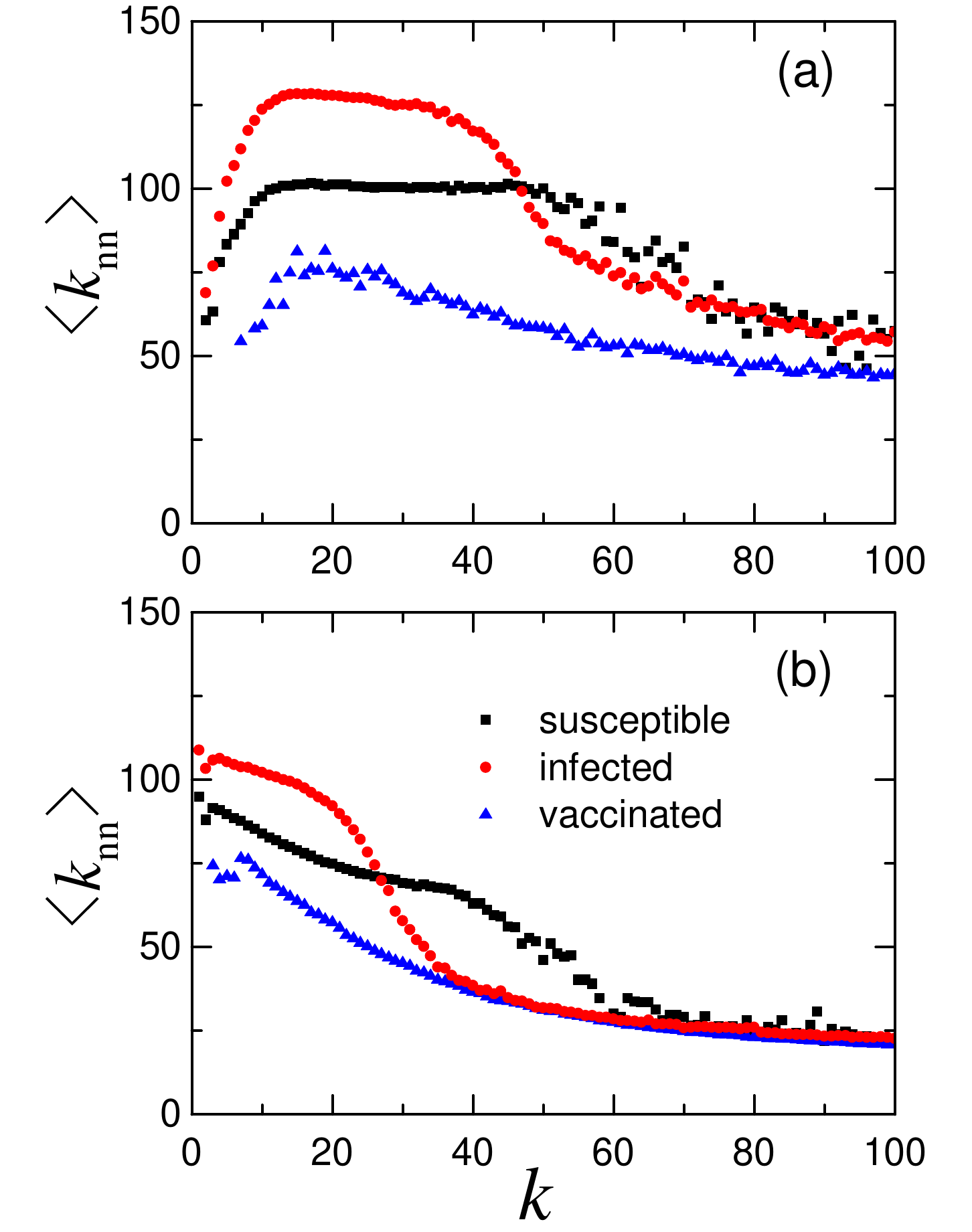}
\caption{(Colour online) Mean nearest-neighbour degree $\langle k_{nn}\rangle$ as a function of any given node degree $k$ in each class of nodes for different epidemic prevalences in the bistable endemic states [Fig.~\ref{fig2}(b)]: a high
epidemic prevalence $i=0.95$ (a), and a low epidemic prevalence $i=0.53$ (b). The black boxes, red circles, and blue triangles represent susceptible, infected, and vaccinated nodes, respectively. Parameters are $N=10^4$, $E=10^5$,
$\alpha=0.006$, $\beta=0.002$, $\varphi=0.00008$, $\psi=0.0002$, $\delta=0.0002$, and $\omega=0.04$.} \label{fig6}\end{center}
\end{figure}

Another commonly studied topological property of networks is the degree correlation (or mixing pattern), which can be
characterised by measuring the average degree of the nearest neighbours of nodes with degree $k$ \citep{PVV01},
\begin{equation}\label{eq:37}
\langle k_{nn}\rangle=\sum_{k'}k'P(k'|k).
\end{equation}
Here, $P(k'|k)$ is the conditional probability that a link belonging to a node with degree $k$ points to a node with degree $k'$. There is a positive (negative) degree correlation if $\langle k_{nn}\rangle$ increases (decreases) with $k$, corresponding to the assortative (disassortative) mixing pattern, where high-degree nodes preferentially connect to highly (sparsely) connected ones \citep{N02}. Otherwise, the network is degree uncorrelated as $k_{nn}$ is independent of $k$. Similar to the work by \cite{GDB06}, the original network is set to be uncorrelated where the initial configuration displays no degree correlations. After we incorporate rewiring into the networked SIV model \citep{PXFZ13}, substantial degree correlations appear. Figure~\ref{fig6} shows the average degree of the nearest neighbours $\langle k_{nn}\rangle$ of degree-$k$ nodes as a function of $k$ for each class of nodes. In the case of a high prevalence $i=0.95$ [Fig.~\ref{fig6}(a)], the network shows a positive degree correlation for small degrees. On the contrary, for larger $k$ the network is disassortative mixing. Similarly, in the case of a low prevalence $i=0.53$ [Fig.~\ref{fig6}(b)], the network is generally disassortative mixing as $\langle k_{nn}\rangle$ decreases with degree $k$ and eventually closes to $\langle k\rangle$. The appearance of the disassortative mixing pattern possibly arises from the fact that $IV$-links are not allowed for rewiring and that the vaccinated nodes continue to accumulate connections when susceptible nodes rewire towards them, until the vaccinated nodes are infected or turn to being susceptible again. On the other hand, the rewiring mechanism separates the infected and susceptible nodes, and typically divides the network into an infected cluster and a susceptible cluster. Note that the mean degrees of infected nodes and susceptible nodes are both much smaller than the mean degree of vaccinated nodes. Therefore, the original susceptible (or infected) nodes that have relatively small degrees are connected to the new susceptible (or infected) nodes that are transformed from vaccinated nodes with high degrees, giving rise to a negative degree correlation.

\subsection{The model with demography}

The adaptive SIV model with demographic effects is expressed as a set of Eqs.~(\ref{eq:10})--(\ref{eq:18}), which can be integrated with a standard numerical integration scheme. In this paper, the
fourth-order Runge-Kutta iteration algorithm is employed to obtain the numerical solution of the model. The set of ordinary differential equations is of dimension $O(k_{\rm {max}})$, where $k_{\rm {max}}$ is the maximal degree value among all nodes. With Eqs.~(\ref{eq:10})--(\ref{eq:18}) and notation in Tables \ref{table1} and \ref{table2}, we can explore the temporal evolution of both the epidemic and the underlying network structure. In particular, we derive not only the time series for the number of infectious nodes and the number of \textit{SI}-links, but also the degree distributions for each class of individuals in the steady state.

We confirm the results derived from the above equations by explicitly simulating the dynamics over two different initial networks --- ER \citep{ER59} and SF networks. The ER random network with a Poisson degree distribution is constructed by linking together each pair of nodes at a certain probability, while an uncorrelated SF random network  is generated following the \textit{uncorrelated configuration model} proposed by \citet{CBP05}. We start the system dynamics by initializing a small fraction $\epsilon$ ($\epsilon=10^{-4}$ in this paper) of infectious seeds. We also initially vaccinate a fraction $\epsilon$ of nodes to ensure the denominator of the fraction ${g^{\prime\prime}_{V}(1,0)}/{g^{\prime}_{V}(1,0)}$ is not equal to zero at the beginning of the numerical calculation. Hence, the initial conditions for the whole system of Eqs.~(\ref{eq:10})--(\ref{eq:18}) are given by:
\begin{eqnarray}
N_{I}(0)&=&N_{V}(0)=\epsilon N(0); \quad
N_{S}(0)=(1-2\epsilon) N(0),\nonumber\\
I_{k}(0)&=&V_{k}(0)=\epsilon p_k(0)N(0); \quad
S_{k}(0)=(1-2\epsilon p_k(0))N(0),\nonumber\\
g^{\prime}_{S}(1,0)&=&\sum_{k}\frac{kS_{k}(0)}{N_S(0)}; \quad
g^{\prime}_{I}(1,0)=\sum_{k}\frac{kI_{k}(0)}{N_I(0)},\nonumber\\
g^{\prime}_{V}(1,0)&=&\sum_{k}\frac{kV_{k}(0)}{N_V(0)}; \quad
M_{S}(0)= g^{\prime}_{S}(1,0)N_{S},\nonumber\\
M_{I}(0)&=&g^{\prime}_{I}(1,0)N_{I}; \quad
M_{V}(0) = g^{\prime}_{V}(1,0)N_{V},\nonumber
\end{eqnarray}
and for a small enough $\epsilon$ ($\epsilon\ll 1$),
\begin{eqnarray}
M_{SI}(0)&=&M_{I}(0);\quad
M_{SV}(0)=M_{V}(0),\nonumber\\
M_{SS}(0)&=&\frac{M_S(0)-M_{SI}(0)-M_{SV}(0)}{2},\nonumber\\
M_{II}(0)&=&M_{IV}(0)=M_{VV}(0)=0.\nonumber
\end{eqnarray}
Henceforth, the parameter values are set to $\alpha=0.06$, $\beta=0.02$, $\varphi=0.0008$, $\psi=0.002$, $\delta=0.0002$, $\omega=0.01$, and $N=10^4$, unless otherwise stated. The degree distribution of newborns is assumed to be the same as the initial network, i.e., $\bar{p}_k=p_k(0)$. For ER networks, we use a Poisson distribution $p_k(0)={e^{-\langle k\rangle}\langle k\rangle ^ {k}}/{k!}$, with $\langle k\rangle=3$. For random SF networks, we choose a power-law distribution $p_k(0)= Ck^{-3}e^{-k/\kappa_{\rm c}} (k_{\rm min}\leq k\leq k_{\rm max})$ with an exponential cutoff $\kappa_{\rm c}$, where $C$ is the normalisation constant, $k_{\rm min}$ is the minimal degree and $k_{\rm max}$ is the maximal degree. In our work we set them to be $\kappa_{\rm c}=80$, $k_{\rm min}=2$, and $k_{\rm max}=30$ to ensure that the initial average degree is $\langle k\rangle\simeq3$. The procedures of the computational simulation at each time step include transmission, rewiring, birth and death events, respectively.

\begin{figure}
\begin{center}
\includegraphics[width=\columnwidth]{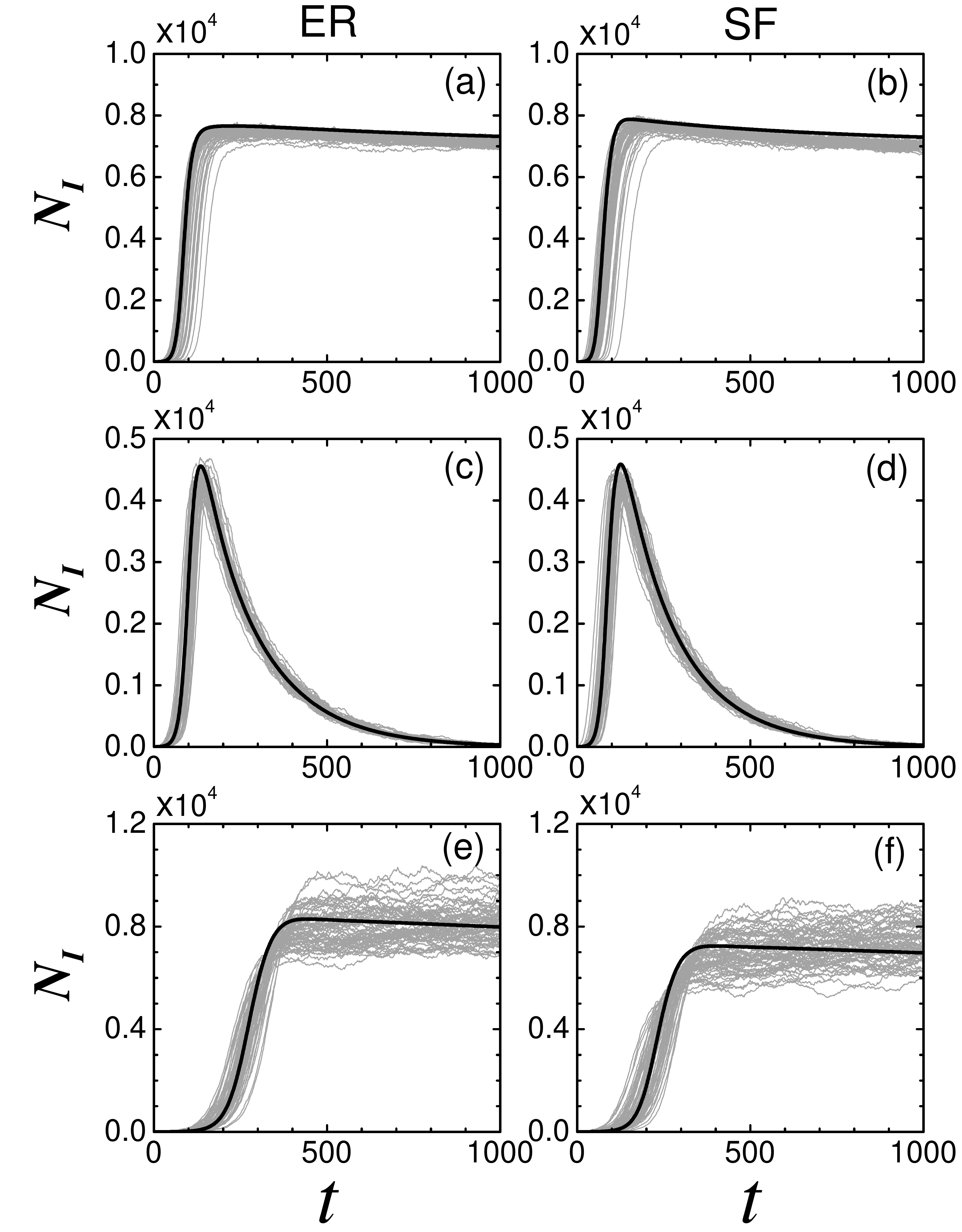}
\caption{Temporal evolution of the number $N_I$ of infected nodes for different starting networks: ER random network [on the left column, i.e., (a), (c), and (e)] and random SF network [on the right column, i.e., (b), (d), and (f)], with the same average degree $\langle k\rangle=3$. The black solid (thick) lines represent the analytical results obtained from the Eqs.~(\ref{a.1}), while the gray dashed (thin) lines correspond to $50$ independent computational simulations. The results are obtained for different parameter values: without birth and death, $\eta_1=\eta_2=\mu=0, \alpha=0.06$ [see (a) and (b)]; with birth and death, $\eta_1=\eta_2=\mu=0.01, \alpha=0.06$ [see (c) and (d)]; with birth and death, $\eta_1=0.01335, \eta_2=\mu=0.01, \alpha=0.03$ [see (e) and (f)]. The other parameters are common: $\beta=0.02$, $\varphi=0.0008$, $\psi=0.002$, $\delta$=0.0002, $\omega=0.01$, and $N=10^4$.}\label{fig7}\end{center}
\end{figure}

\subsubsection{Epidemic patterns}

The temporal evolution of the number of infected nodes $N_I$ is presented in Fig.~\ref{fig7}, exhibiting a good agreement between the simulation results (gray dashed lines) and the analytical solutions (black solid lines) obtained from the model equations for different parameter values. Figures~\ref{fig7}(a) and \ref{fig7}(b) show the time series of the number of infectives, respectively, in the ER and SF networks with $\langle k\rangle=3$, where the birth and death processes are not allowed ($\eta_1=\eta_2=\mu=0$). Due to the rewiring mechanism, the analytical trajectory shows a very similar trend for both networks; however, the simulation trajectories exhibit significant variability in terms of the time required to reach the expansion phase and the epidemic prevalence. This is mainly due to the stochastic effect of early infections. An initially infected low-degree node, which takes much time to infect its neighbours, will delay the onset of the expansion phase, while the initial infection of a high-degree node may lead to an explosive outbreak \citep{Volz08}. On the contrary, if we introduce nonzero rates of birth and death $\eta_1=\eta_2=\mu=0.01$, the disease will eventually die out after an initial outbreak in both ER and SF networks, as shown in Figs.~\ref{fig7}(c) and \ref{fig7}(d). This happens because the inadequate birth rate fails to offset the natural and disease-induced deaths. Again, the trajectories of infected nodes in both networks are almost the same. Finally, Figs.~\ref{fig7}(e) and \ref{fig7}(f) show a series of independent simulations for the number of infectives at a larger birth rate $\eta_1=0.01335$ and a smaller transmission rate $\alpha=0.03$ in the ER and SF networks, respectively. In this case, the analytical trajectory traverses through the central region of the dense swarm of simulation trajectories, and the black thick lines in Figs.~\ref{fig7}(e) and \ref{fig7}(f) demonstrate that on average the epidemic prevalence in the SF network is smaller than that in the ER network (particularly at $t=1000$). The possible reason for this disparity is rooted in the heterogeneity among the node degrees. In the SF network high-degree nodes are more likely to get infected and hence they die more often (due to a more rapid death), which in turn leads to a greater reduction in the average degree of the entire network and consequently results in a lower infection level. Compared with Figs.~\ref{fig7}(a) to \ref{fig7}(d), the simulation trajectories in Figs.~\ref{fig7}(e) and \ref{fig7}(f) exhibit stronger variability in the epidemic prevalence. Since the birth rate is greater than the natural death rate, the continuous input of newborn nodes will lead to significant changes in the total network size and the connectivity pattern of the entire network during the epidemic outbreak, and hence the epidemic prevalence.

\begin{figure}
\begin{center}
\includegraphics[width=\columnwidth]{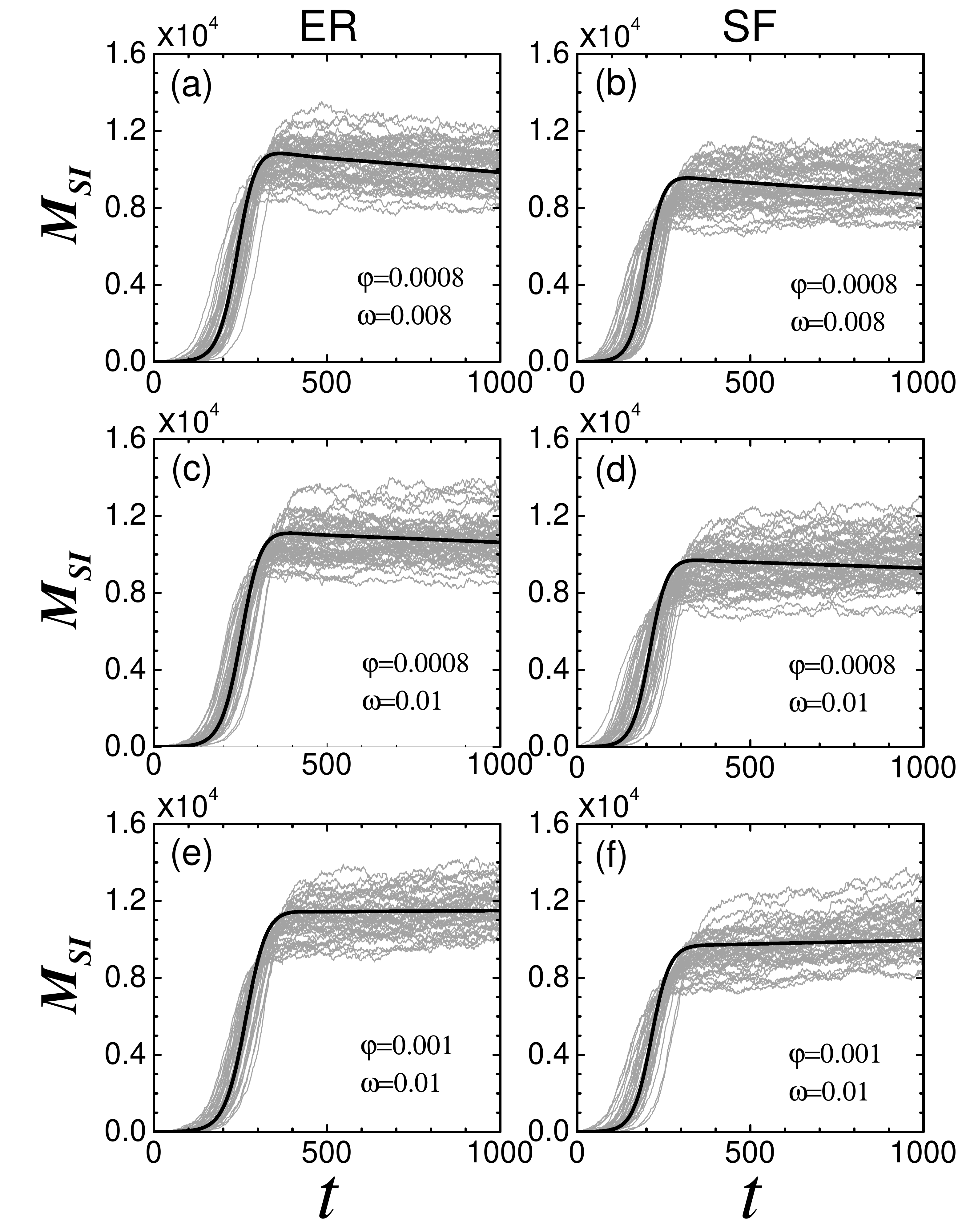}
\caption{Temporal evolution of the number of \textit{SI}-links in ER networks [on the left side: (a), (c), and (e)] and SF networks [on the right side: (b), (d), and (f)], with the same average connectivity $\langle k\rangle=3$. The black (thick) lines indicate the theoretical results obtained by numerical integration of the Eq.~(\ref{eq:14}), and the gray (thin) lines are the simulation results of at least 50 independent realizations. The results are obtained by varying the vaccination rate $\varphi$ and the rewiring rate $\omega$ [here, we choose $\varphi=0.0008, \omega=0.008$ for (a) and (b); $\varphi=0.0008, \omega=0.01$ for (c) and (d); $\varphi=0.001, \omega=0.01$ for (e) and (f)], while the other parameters are the same as in Figs.~\ref{fig7}(e) and \ref{fig7}(f): $\eta_1=0.01335, \eta_2=\mu=0.01, \alpha=0.03, \beta=0.02, \psi=0.002, \delta=0.0002$, and $N=10^4$.}\label{fig8}\end{center}
\end{figure}

\begin{figure}
\begin{center}
\includegraphics[width=\columnwidth]{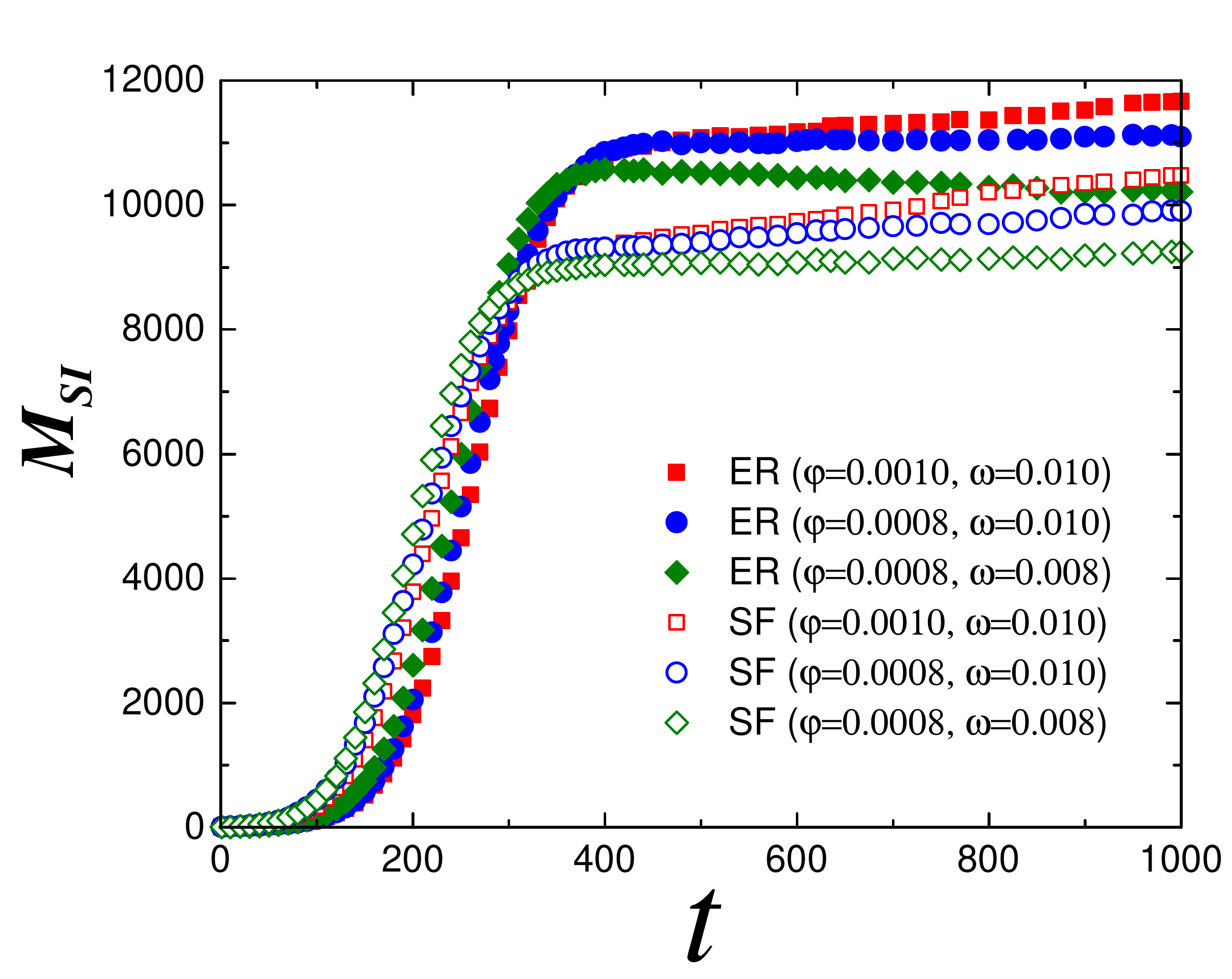}
\caption{(Colour online) Temporal evolution of the number of \textit{SI}-links in ER networks (filled symbols) and SF networks (empty symbols). All the symbols are the simulation results averaged over at least 50 independent realizations. The results are obtained by varying the vaccination rate $\varphi$ and the rewiring rate $\omega$, as indicated in the legend. The other parameters are: $\eta_1=0.01335, \eta_2=\mu=0.01, \alpha=0.03, \beta=0.02, \psi=0.002, \delta=0.0002$, and $N=10^4$.}\label{fig9}\end{center}
\end{figure}

\begin{figure}
\begin{center}
\includegraphics[width=\columnwidth]{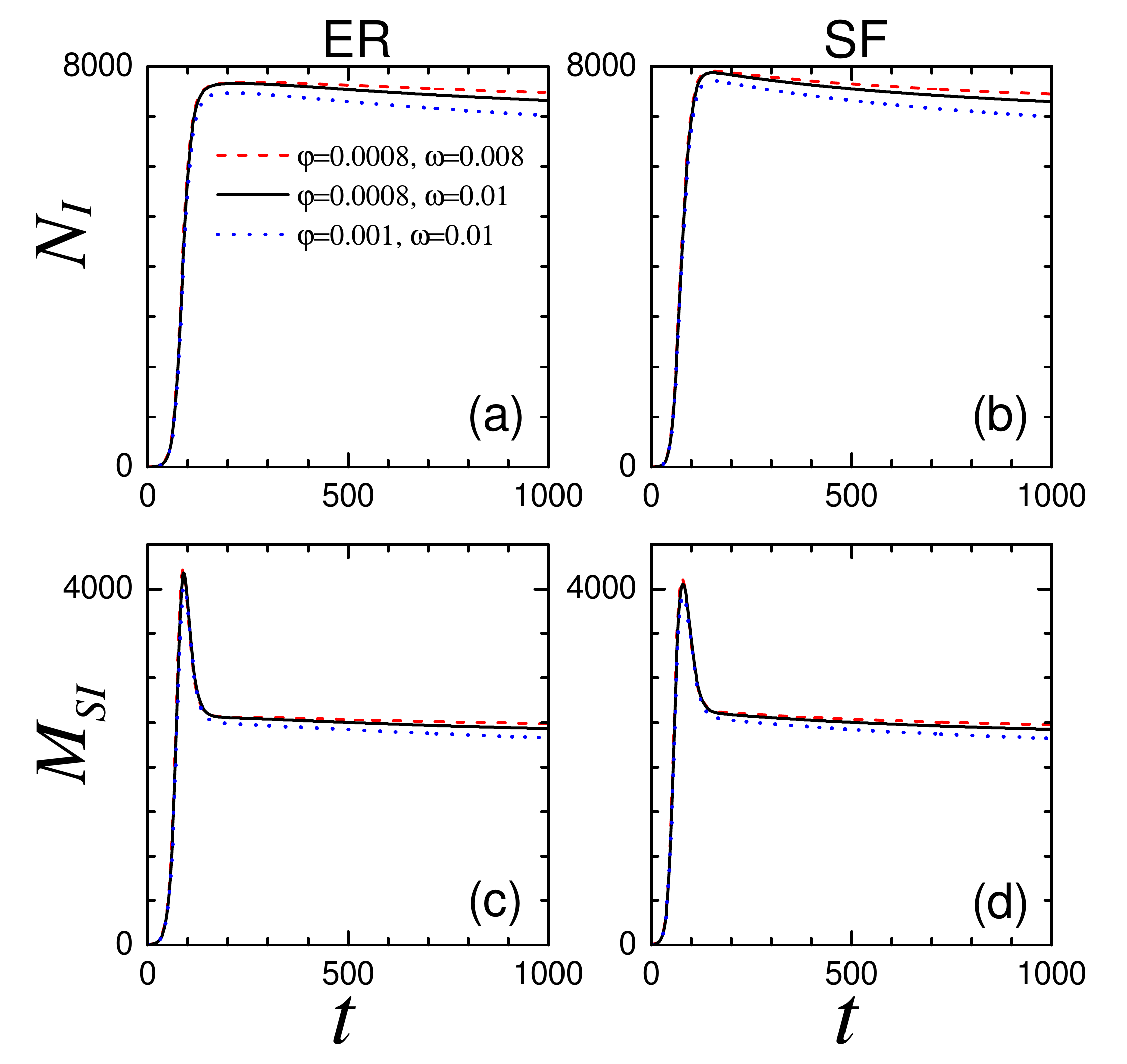}
\caption{(Colour online) Temporal evolution of $N_I$ (top) and $M_{SI}$ (bottom) in ER networks [on the left side: (a) and (c)] and SF networks [on the right side: (b) and (d)] for the adaptive SIV model without birth and death processes. In this case only the theoretical results are presented. The red dashed lines indicate the results for $\varphi=0.0008$ and $\omega=0.008$; the black solid lines for $\varphi=0.0008$ and $\omega=0.01$; and the blue dotted lines for $\varphi=0.001$ and $\omega=0.01$. Other parameters are $\alpha=0.06$, $\beta=0.02$, $\psi=0.002$, $\delta=0.0002$, $\eta_1=\eta_2=\mu=0$.}\label{fig10}
\end{center}
\end{figure}

To understand in more detail how the epidemic spreading is influenced by the adaptive rewiring behavior of susceptible individuals and the random vaccination process, we further investigate the temporal evolution of the number of contacts between susceptible and infected nodes, as shown in Fig.~\ref{fig8}. For the parameter values ($\varphi=0.0008$ and $\omega=0.01$) given in Figs.~\ref{fig7}(e) and~\ref{fig7}(f), the number of \textit{SI}-links, $M_{SI}$, in both ER [Fig.~\ref{fig8}(c)] and SF [Fig.~\ref{fig8}(d)] networks is shown to have the same qualitative behavior as the number of infected nodes, $N_I$. The \textit{SI}-links are the major infectious contacts since the probability $\delta\alpha$ for vaccinated individuals to be infected is extremely limited ($\delta=0.0002$). In this regard, the number $M_{SI}$ typically determines the number $N_I$. To show a clear comparison between the effects of public vaccination and adaptive rewiring on the number of $SI$-links, we present the average simulation results in Fig.~\ref{fig9}, in which the number $M_{SI}$ of $SI$-links in the ER network increases as the rewiring rate is raised from $\omega=0.008$ to $\omega=0.01$ with the vaccination rate remaining the same ($\varphi=0.0008$). Still, the number $M_{SI}$ is further enlarged as $\varphi$ is increased from $\varphi=0.0008$ to $\varphi=0.001$, with $\omega$ keeping unchanged ($\omega=0.01$). This implies that as the parameters $\varphi$ and $\omega$ are both increased from $\varphi=0.0008$, $\omega=0.008$ to $\varphi=0.001$, $\omega=0.01$, the increment of the number $M_{SI}$ is larger than the increment of $M_{SI}$ caused by increasing either $\varphi$ or $\omega$ alone. In this regard, we conclude that public vaccination, in conjunction with adaptive rewiring, facilitates disease spread. In combination with the birth and death processes, lowering the rewiring rate causes $M_{SI}$ to decrease, whereas enhancing the vaccination rate increases $M_{SI}$. This case is in sharp contrast to the model without demographic processes, in which both the rewiring and vaccination contribute to decrease $M_{SI}$ and hence $N_I$, as shown in Fig.~\ref{fig10}. These counterintuitive results imply that in the case of a varying population with demographic effects (births, deaths and migration), susceptible individuals' protective behavior and the public vaccination interventions that aim to suppress the epidemic, would make the situation even worse. These phenomena can be explained as follows. Intuitively, infected nodes die at a more rapid rate since $\mu>0$. When the rewiring rate is increased, the infected nodes have lower degree when they die. Therefore, the average degree of the entire network is not reduced as much by node death as it would be in the absence of rewiring. Higher mean degree could lead to more infections. When the vaccination rate is increased, then there are more vaccinated nodes (that are able to accumulate connections from rewiring and thus have high degrees) serving as the hub nodes in the network. On the other hand, the adaptive rewiring of $SI$-links, in conjunction with the birth and death processes, contributes to a structural change in the network. Typically, rewiring tends to increase clustering --- $SI$-links are broken and then susceptible nodes are forced to select new neighbours from among the more limited subset of noninfected nodes (which will be in a topologically constrained part of the network). Imagine an infected subgraph, the susceptible nodes at the boundary are rewiring their connection to other noninfected nodes elsewhere increasing clustering. Clustering (particularly with negative correlation between degrees) means that once infection reaches these regions (and the hub nodes in particular) one would expect worse outbreak. As a matter of fact, infection is facilitated in reaching these regions with the addition of new nodes via the birth and death processes (since they are connected randomly across the existing structure). Moreover, it is shown from Fig.~\ref{fig9} that the number $M_{SI}$ in SF networks expands faster but is more restricted than in ER networks. The possible reason for this difference is rooted in the heterogeneity among the node degrees. In the SF network high-degree nodes are more likely to get infected and hence they die more often (due to a more rapid death rate), which in turn leads to a greater reduction in the average degree of the entire network and consequently results in a lower infection level.

\begin{figure}
\begin{center}
\includegraphics[width=\columnwidth]{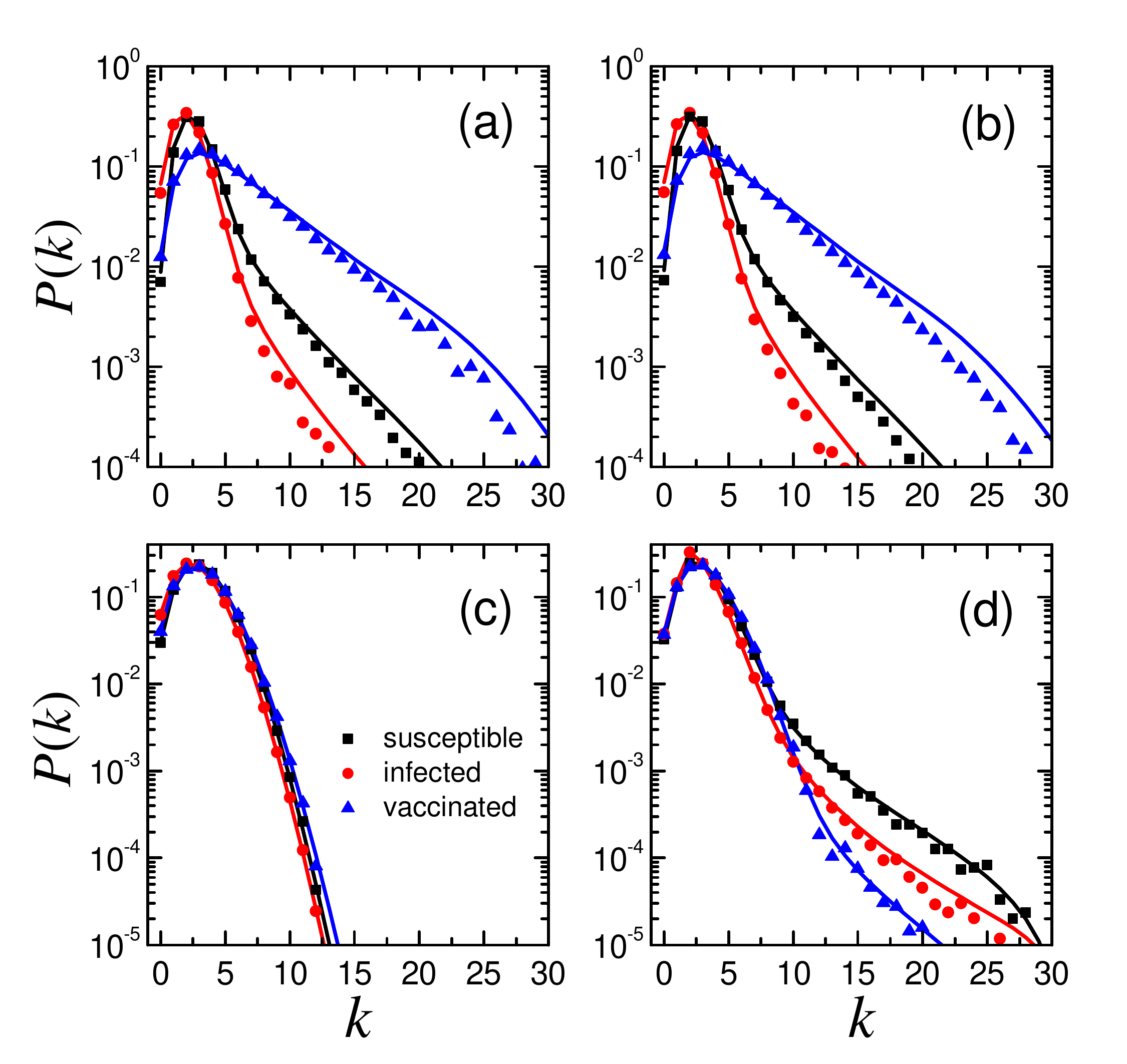}
\caption{(Colour online) Degree distributions of nodes in each class in the stationary state at $t=1000$. The results on the left (right) column are obtained from an initial ER (SF) network, with an average degree $\langle k\rangle=3$. The black boxes, red circles and blue triangles stand for susceptible, infected, and vaccinated individuals, respectively. The symbols correspond to simulation results averaged over 100 different realizations and the solid lines are numerical solutions obtained from Eqs.~(\ref{eq:10})--(\ref{eq:12}) and (\ref{a.1}). The parameters in the panels (a) and (b) are the same as in Figs.~\ref{fig7}(a) and \ref{fig7}(b), while those in the panels (c) and (d) are the same as in Figs.~\ref{fig7}(e) and \ref{fig7}(f).}\label{fig11}\end{center}
\end{figure}

\begin{figure}
\begin{center}
\includegraphics[width=\columnwidth]{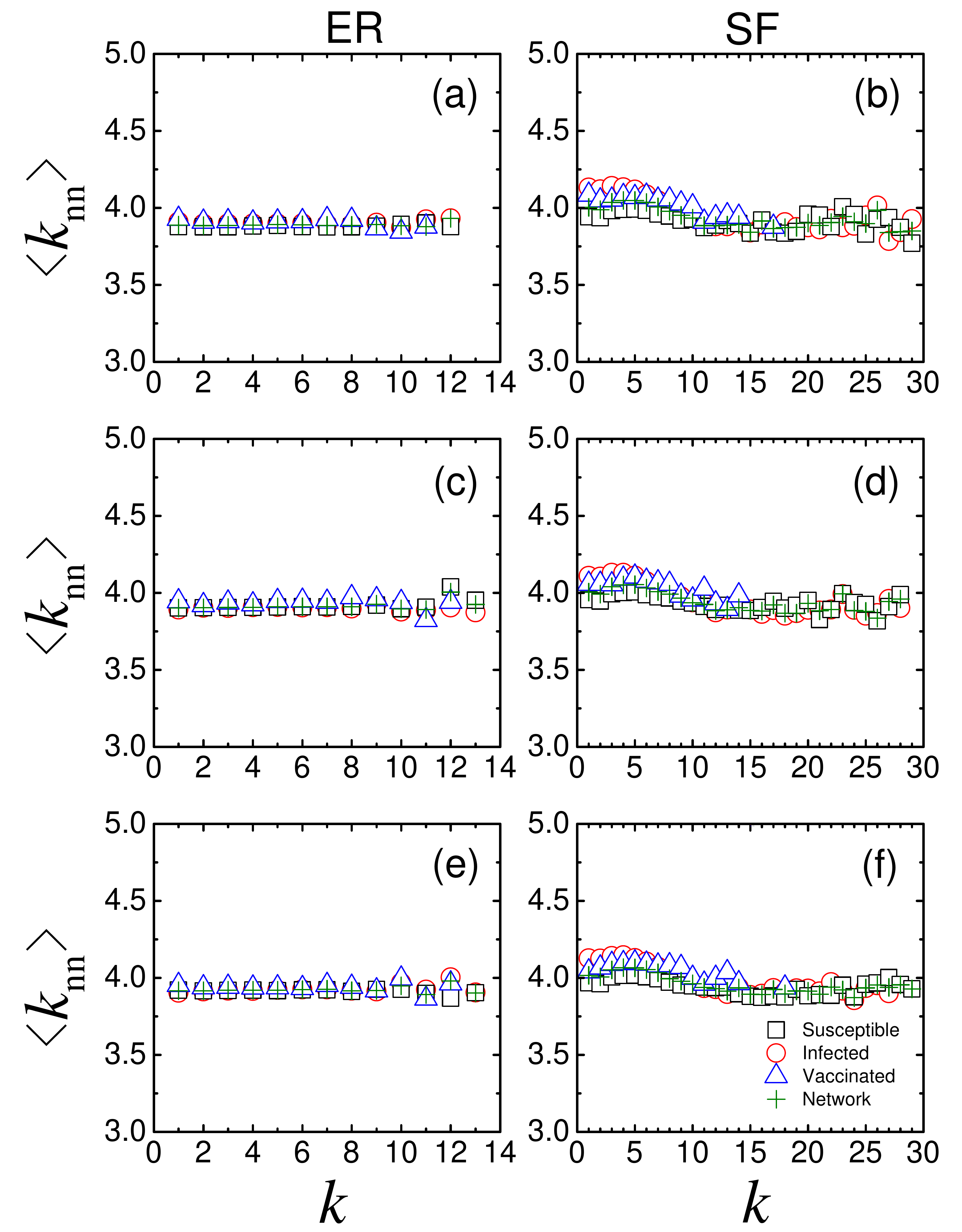}
\caption{(Colour online) Mean nearest-neighbour degree $\langle k_{\rm {nn}}\rangle$ as a
function of any given degree $k$ in each class of nodes in the state at $t=1000$. The results on the left (right) column are obtained from an initial ER (SF) network, with an average degree $\langle k\rangle=3$. The empty boxes (black) ``$\Box$", circles (red) ``$\bigcirc$" and triangles (blue) ``$\triangle$" represent susceptible, infected and vaccinated nodes, respectively, and the crosses (olive) ``$+$" denote the results of the entire network. All these symbols are simulation results obtained by averaging over 100 different realizations. The corresponding parameter values in each diagram are the same as in each diagram of Fig.~\ref{fig8}.}\label{fig12}\end{center}
\end{figure}

\subsubsection{Network topology}

Furthermore, our model is able to capture the topological structure of the network in the stationary state of the epidemic.  In particular, the degree distribution in each class of nodes is shown in Fig.~\ref{fig11}, where the black, red, and blue colours correspond to the susceptible, infected, and vaccinated nodes, respectively. To be precise, the symbols represent the simulation results and the solid lines are the analytical solutions to system equations. As shown in Figs.~\ref{fig11}(a) and \ref{fig11}(b), without the birth and death processes, the network evolves to the same degree distribution due to the rewiring mechanism. On the contrary, the situation is quite different with the introduction of birth and death processes. For instance, in Figs.~\ref{fig11}(c) and \ref{fig11}(d) we exhibit the degree distribution of each class of nodes at the final time step $t=1000$, for the settings in Figs.~\ref{fig7}(e) and \ref{fig7}(f), respectively. In the presence of node birth, different degree distributions of the newborns will lead to different network structures. It is clear that the final network is narrowly distributed if nodes are born with a poisson degree distribution [Fig.~\ref{fig11}(c)], while the network evolves to a broader degree distribution if the newborns follow a power-law degree distribution [Fig.~\ref{fig11}(d)]. It is worth noting that by the mathematical model of Eqs.~(\ref{eq:10})--(\ref{eq:18}), the theoretical results of degree distribution can be obtained not only for the model with demography, but also for the case without demography where the birth and death rates are set to zero, as illustrated in the Figs.~\ref{fig11}(a) and \ref{fig11}(b). Nevertheless, under the modelling framework of Eqs.~(\ref{eq:1})--(\ref{eq:9}) where network properties are not included, the theoretical results of degree distribution cannot be derived.

As mentioned above, since the birth process dominates in affecting the degree distribution of nodes, the network structure is degree uncorrelated, which is shown in Fig.~\ref{fig12}. Although the individual rewiring changes the local connectivity pattern among nodes, the continuous input of the newborns that randomly link to old ones breaks the degree correlation that emerges in the scenario without birth and death processes (Fig.~\ref{fig6}). Note that the average nearest-neighbour degree is close to $4$, $\langle k_{\rm {\rm {nn}}}\rangle\simeq 4$, which is larger than the average degree of the network $\langle k\rangle$, which is around $2.9$ in our simulations. This is consistent with the friendship paradox. Similar results are observed for different parameter values among each class of nodes in both ER and SF networks (Fig.~\ref{fig12}). It is worth noting that although newborn nodes arrive with mean degree $3$, it will not still be the mean degree of the network due to a more rapid death of infectives. Actually, the mean degree of the entire network is parameter dependent.

\section{Conclusions}
\label{sec4}

Widespread public vaccination programmes and individual protective behavior are two key factors of disease control. Ignoring either component will reduce the accuracy of our models or, ultimately, the effectiveness of containment measures. In this paper, we developed a mathematical framework to model the combined effects of public vaccination intervention and individual protective behavior on epidemic dynamics by studying an SIS model with imperfect vaccination in a dynamic contact network. The model was analyzed in a closed population without demographic turnover and a varying population with births, deaths and migration.

In the case of a closed population (without demographic effects), we focused specifically on the combined effects of macroscopic random vaccination and microscopic contact rewiring on the system dynamics by conducting a bifurcation analysis of the model. Our analysis shows that compared with either individual strategy, the combination of vaccination programme and adaptive rewiring contributes to a more effective control for the spread of infectious diseases by increasing both the invasion threshold and the persistence threshold of the epidemic. The larger the vaccination and rewiring rates are, the safer the population is from infectious diseases. The occurrence of stable epidemic cycle at a low prevalence is particularly relevant for the seasonality of influenza epidemics \citep{Earn02}. Moreover, the conjunction of vaccination and rewiring has significant impacts on the topology of the underlying contact network. Both the degree distributions and degree correlations among different classes of nodes are characteristically distinct.

In the case of a varying population with demographic development (births, deaths and migration), we studied the epidemic patterns and the network topology on the basis of the dynamical behavior of nodes and links at a microscopic level, using both the probability generating function and the pairwise approximation. This modelling approach allows us to incorporate the effects of vaccination and rewiring as well as the demographic changes into network-based epidemiological models with arbitrary degree distributions. The analytical results of the temporal evolution of both the number of infective individuals and the number of infectious contacts, as well as the degree distribution of nodes in each class, have shown a good agreement with extensive computational simulations. In general, diseases in SF networks spread relatively faster yet end at a lower level of epidemic prevalence, as compared with the ER networks. Ironically, the combination of contact rewiring and random vaccination aiming to contain the epidemic would probably facilitate the disease spread. While the network without demography evolves to an identical topology for different initial network configurations, the network with demography develops into different structures depending on the degree distribution of the newborns. Besides, the underlying network structure displays no degree correlation regardless of model parameters and initial settings.

In summary, the present work manifests the significant combined effects of public vaccination intervention and individual protective behavior on both the epidemic dynamics and the underlying network structure. Moreover, the strikingly different results for the two cases of population assumptions indicate the strong impacts of the demographic turnover. Thus, this research provides a more comprehensive insight into disease containment strategies.

\begin{acknowledgements}
The authors would like to thank the two anonymous referees for their valuable comments and suggestions which have greatly improved this paper. This work was supported by NSFC grants (11331009, 11501340) and STCSM grant (13ZR1416800). MS was supported by an Australian Research Council Future Fellowship (FT 110100896) and Discovery Project (DP 140100203).
\end{acknowledgements}

\section*{Appendix A: Total number and degree of nodes in each class}\label{appendix.a}
\setcounter{equation}{0}
\renewcommand{\theequation}{A\arabic{equation}}

Summing Eqs.~(\ref{eq:10})--(\ref{eq:12}) over all $k$ values leads to the differential equations for the total number of susceptibles $N_{S}$, infecteds $N_{I}$, and vaccinateds $N_{V}$ as follows:
\begin{equation}\label{a.1}
\left.\begin{array}{ccl}
\frac{\rm d}{{\rm d}t}N_{S} = \sum_{k}\frac{\rm d}{{\rm d}t}S_{k} &=& \eta_{1}N+\beta N_{I}+\psi N_{V}\\
&& -(\eta_{2}+\varphi)N_{S}-\alpha M_{SI},\\
&&\\
\frac{\rm d}{{\rm d}t}N_{I} = \sum_{k}\frac{\rm d}{{\rm d}t}I_{k} &=& \alpha M_{SI}+\delta\alpha M_{IV}\\
&&-(\eta_{2}+\mu+\beta)N_{I},\\
&&\\
\frac{\rm d}{{\rm d}t}N_{V} = \sum_{k}\frac{\rm d}{{\rm d}t}V_{K} &=& \varphi N_{S}-(\eta_{2}+\psi)N_{V}\\
&&-\delta\alpha M_{IV}.
\end{array}\right\}
\end{equation}
As a consequence, the total network size $N=N_S+N_I+N_V$ varies in time such that
\begin{equation}\label{a.2}
  \frac{\rm d}{{\rm d}t}N=(\eta_1-\eta_2)N-\mu N_I,
\end{equation}
which reflects the demographic change. Furthermore, the dynamical equations for the total degrees of nodes in each state can be derived as:
\begin{equation}\label{a.3}
\left.\begin{array}{lll}
\frac{\rm d}{{\rm d}t}M_{S}&=& \sum_{k}k(\frac{\rm d}{{\rm d}t}S_{k})\\
 &=& \eta_{1}\bar{g}^{\prime}(1)(N+N_{S})+\beta M_{I}
+\psi M_{V}+\omega M_{SI}\frac{N_{S}}{N_{S}+N_{V}}\\
&&-(2\eta_{2}+\varphi)M_{S}-\mu M_{SI}
-\alpha\frac{M_{SI}}{M_{S}}N_{S}[g^\prime_{S}(1,t)+g^{\prime\prime}_{S}(1,t)],\\
\frac{\rm d}{{\rm d}t}M_{I}&=&\sum_{k}k(\frac{\rm d}{{\rm d}t}I_{k}) \\
&=&\eta_{1}\bar{g}^{\prime}(1)N_{I}-2\mu M_{II}-\omega M_{SI}
+\alpha\frac{M_{SI}}{M_{S}}N_{S}[g^{\prime}_{S}(1,t)
+g^{\prime\prime}_{S}(1,t)]\\
&&+\delta\alpha\frac{M_{IV}}{M_{V}}N_{V}[g^{\prime}_{V}(1,t)
+g^{\prime\prime}_{V}(1,t)]-(2\eta_{2}+\mu+\beta)M_{I},\\
\frac{\rm d}{{\rm d}t}M_{V}&=&\sum_{k}k(\frac{\rm d}{{\rm d}t}V_{k})\\
&=&
\eta_{1}\bar{g}^{\prime}(1)N_{V}+\omega M_{SI}\frac{N_{V}}{N_{S}+N_{V}}
 +\varphi M_{S}-(2\eta_{2}+\psi)M_{V}\\
 &&-\mu M_{IV}-\delta\alpha\frac{M_{IV}}{M_{V}}N_{V}[g^{\prime}_{V}(1,t)+g^{\prime\prime}_V(1,t)],
\end{array}\right\}
\end{equation}
which yield the evolution equation for the total number $E$ of links
\begin{equation}\label{a.4}
  \frac{\rm d}{{\rm d}t}E=\frac{\rm d}{{\rm d}t}\frac{\sum_{A}M_A}{2}=\eta_1 \bar{g}^{\prime}(1)N-2\eta_2 E-\mu M_I.
\end{equation}
Note that $g^{\prime}_{A}$ and $g^{\prime\prime}_{A}$ are the first-order and the second-order derivatives of the generating function $g_{A}(x,t)$, respectively, with regard to variable $x$, where $A\in \{S, I, V\}$ (see also Table \ref{table2} ). In particular, one gets
\begin{equation}\label{a.5}
\left.\begin{array}{rll}
g_{S}^{\prime}(1,t)=\sum_{k}kp_{Sk}&=&\sum_{k}\frac{kS_{k}}{N_S},\\
&&\\
g_{V}^{\prime}(1,t)=\sum_{k}kp_{Vk}&=&\sum_{k}\frac{kV_{k}}{N_V},\\
&&\\
g_{S}^{\prime\prime}(1,t)=\sum_{k}k(k-1)p_{Sk}&=&\sum_{k}\frac{k(k-1)S_{k}}{N_S},\\
&&\\
g_{V}^{\prime\prime}(1,t)=\sum_{k}k(k-1)p_{Vk}&=&\sum_{k}\frac{k(k-1)V_{k}}{N_V}.
\end{array}\right\}
\end{equation}

\section*{Appendix B: Approximation for the number of triples}\label{appendix.b}
\setcounter{equation}{0}
\renewcommand{\theequation}{B\arabic{equation}}

Employing the technique of generating functions used in \citet{HK11}, we generalise the standard moment closure approximation in \citet{Keeling97} for the number of triples accountable for new infections to incorporate the heterogeneity in the central node's degree. For example, consider an $SI$-link connecting an infective with a susceptible, the probability for the susceptible to have degree $k$ is ${kS_{k}}/{(\sum_{k'}k'S_{k'})}$, and among the remaining $(k-1)$ neighbours, there is a fraction
${M_{SI}}/{M_S}$ of infectious individuals. Here, we assume that the probability for a link starting from a susceptible to point to an infective is independent of the starting node's degree $k$. Therefore, from the mean-field perspective the average number $M_{ISI}$ of $ISI$-type triples is
\begin{eqnarray}
M_{ISI}&=&M_{SI}\sum_{k}\frac{kS_{k}}{\sum_{k'}k'S_{k'}}(k-1)\frac{M_{SI}}
{M_{S}}\nonumber\\
&=& M_{SI}\frac{M_{SI}}{M_{S}}\frac{g_{S}^{\prime\prime}(1,t)}
{g_{S}^{\prime}(1,t)}.\label{b.1}
\end{eqnarray}
Analogously, one can get the number $M_{SSI}$ of $SSI$-type triples
\begin{eqnarray}
M_{SSI}&=&M_{SI}\sum_{k}\frac{kS_{k}}{\sum_{k'}k'S_{k'}}(k-1)\frac{2M_{SS}}{M_S}\nonumber\\
&=& M_{SI}\frac{2M_{SS}}{M_S}\frac{g_{S}^{\prime\prime}(1,t)}
{g_{S}^{\prime}(1,t)},\label{b.2}
\end{eqnarray}
and the number $M_{ISV}$ of $ISV$-type, $M_{SVI}$ of $SVI$-type, $M_{IVI}$ of $IVI$-type, and $M_{IVV}$ of $IVV$-type triples
\begin{eqnarray}
M_{ISV}&=&M_{SI}\frac{M_{SV}}{M_{S}}\frac{g_{S}^{\prime\prime}(1,t)}
{g_{S}^{\prime}(1,t)},\label{eq:23}\\
M_{SVI}&=&M_{SV}\frac{M_{IV}}{M_{V}}\frac{g_{V}^{\prime\prime}(1,t)}
{g_{V}^{\prime}(1,t)},\label{eq:24}\\
M_{IVI}&=&M_{IV}\frac{M_{IV}}{M_{V}}\frac{g_{V}^{\prime\prime}(1,t)}
{g_{V}^{\prime}(1,t)},\label{eq:25}\\
M_{IVV}&=&M_{IV}\frac{2M_{VV}}{M_{V}}\frac{g_{V}^{\prime\prime}(1,t)}
{g_{V}^{\prime}(1,t)},\label{eq:26}
\end{eqnarray}
where the generating functions are given in (\ref{a.5}).

%




\end{document}